\documentclass[onecolumn]{aastex62}

\usepackage{graphicx}
\usepackage{graphbox}
\usepackage{tikz}
\usepackage{enumitem}
\usepackage{float}
\usepackage{multirow}
\usepackage{color}
\usepackage[colorinlistoftodos]{todonotes}

\accepted{}
\begin{document}
	\title{Impact of planetary mass uncertainties on exoplanet atmospheric retrievals}
	
	\correspondingauthor{Q. Changeat}
	\email{quentin.changeat.18@ucl.ac.uk}
	\author[0000-0001-6516-4493]{Q. Changeat}
	\affil{Department of Physics and Astronomy \\
		University College London \\
		Gower Street,WC1E 6BT London, United Kingdom}
	\author[0000-0001-6058-6654]{L. Keyte}
	\affil{Department of Physics and Astronomy \\
		University College London \\
		Gower Street,WC1E 6BT London, United Kingdom}
	\author[0000-0002-4205-5267]{I.P. Waldmann}
	\affil{Department of Physics and Astronomy \\
		University College London \\
		Gower Street,WC1E 6BT London, United Kingdom}
	\author[0000-0001-6058-6654]{G. Tinetti}
	\affil{Department of Physics and Astronomy \\
		University College London \\
		Gower Street,WC1E 6BT London, United Kingdom}

\begin{abstract}
In current    models used to interpret exoplanet atmospheric observations,  the planetary mass is treated as a prior and is measured/estimated independently with external methods, such as radial velocity or Transit Timing Variation techniques. This approach is necessary as available  spectroscopic data do not have sufficient wavelength coverage and/or signal to noise to infer the planetary mass.    
We examine here whether the planetary mass can be directly retrieved from transit spectra as observed by future space observatories, which will provide higher quality spectra. More in general,  we quantify  the impact of mass uncertainties on spectral retrieval analyses for a host of atmospheric scenarios. 

Our approach is both analytical and numerical: we first  use simple approximations to extract analytically the influence of each atmospheric/planetary parameter to the wavelength-dependent transit depth. We then adopt a fully Bayesian retrieval  model to quantify the propagation of the mass uncertainty onto other atmospheric parameters.

We found that for clear-sky, gaseous  atmospheres  the  posterior distributions are the same when the mass is known or retrieved. The retrieved mass is very accurate, with a precision of more than 10\%, provided the wavelength coverage and S/N are adequate. 
When opaque clouds are included in the simulations, the uncertainties in the retrieved mass increase, especially for high altitude clouds. 
However atmospheric parameters such as the temperature and trace-gas abundances are unaffected by the knowledge  of the mass. 

Secondary atmospheres, expected to be present in many super-Earths, are more challenging due to the higher degree of freedom for the atmospheric main component, which is unknown. For broad wavelength range and adequate signal to noise observations, the mass can still be retrieved accurately and precisely if clouds are not present, and so are all the other atmospheric/planetary parameters.
When clouds are added, we find that the mass uncertainties may impact substantially the retrieval of the mean molecular weight: an independent characterisation of the mass would therefore be helpful to capture/confirm the main atmospheric constituent.

\end{abstract}

\section{INTRODUCTION}

In  recent years, the study of exoplanetary atmospheres has shifted from the investigation of individual planets to the characterisation of populations (e.g. \cite{Barstow_10HJ}, \cite{Tsiaras_pop_study_trans}, \cite{Pinhas_ten_HJ_clouds}). In parallel, detection missions, such as Kepler and TESS, and ground-based observatories are enabling the identification of an increasing number of interesting  targets suitable for atmospheric studies. 
New space  observatories and dedicated missions, such as the NASA James Webb Space Telescope \citep{Bean_JWST} and the ESA ARIEL mission \citep{Tinetti_ariel}, are expected to revolutionise our understanding of the physical and chemical properties of a large and diverse sample of extrasolar worlds.   To prepare for these missions, significant resources will be allocated to acquire/refine basic planetary, orbital and stellar parameters. To maximise the efficiency of the community effort, it is important to prioritise follow-up activities where these are particularly needed. For example, the recent ARIEL ExoClock project\footnote{https://www.exoclock.space} provides  priorities to guide amateur astronomers in the selection of targets for ephemeris refinement. While for most planets considered for transit spectroscopy, the mass is already constrained from  radial velocity observations,  it is important to evaluate whether these measurements are precise enough for atmospheric characterisation. In the case of low gravity exoplanets, current masses may have large uncertainties  and it is therefore foreseen that refinements from radial velocity \citep{L_pez_Morales_2016} or Transit Timing Variation techniques \citep{10.1093/mnras/stz181} will have to be made in preparation for JWST and ARIEL.

Current transit spectroscopic data do not have sufficient wavelength coverage and/or signal to noise to infer the planetary mass \citep{line_info_content}, therefore spectral retrieval models   include this key parameter as a prior estimated through external methods, such as radial velocity or Transit Timing Variation techniques. 
This limitation will no longer apply to  future space missions and observatories (JWST, ARIEL, Twinkle \citep{Edwards_twinkle})  designed to provide spectroscopic observations over a broader wavelength range, higher spectral resolution and signal to noise.  
 \cite{deWit_mass} showed that for atmospheres dominated by a single species, the mass could be retrieved from transit spectra only. \cite{deWit_mass} also stressed the importance of Rayleigh scattering and collision induced absorption, which are particularly valuable to constraint masses from retrievals. However, in their examples, they only considered atmospheres dominated by a single species.
\cite{Batalha_mass} highlighted the degeneracy between mean molecular weight and main atmospheric components for planets with a secondary atmosphere -- i.e. an atmosphere that has evolved from a pure H/He composition. However, they restricted their analysis to the comparison of forward models for the specific case of a H$_2$/H$_2$O atmosphere. 

Here we aim to determine whether the planetary mass can be directly retrieved from transit spectra observed by future space observatories and, more generally, to quantify the impact of mass uncertainties on spectral retrieval analyses. 
Compared to previous studies in the literature, we investigate a more comprehensive suite of atmospheric scenarios and we adopt a fully Bayesian retrieval  to quantify the propagation of the mass uncertainty onto other atmospheric parameters. 
The paper is comprised of two main sections. The first section explores analytically the role of the mass in transit spectroscopy  and  the contribution of the different parameters to the optical depth. In the second section, we use atmospheric retrieval techniques to illustrate/confirm the predictions made in section 1 and estimate the mass uncertainties in various key examples. \\

\section{ANALYTICAL STUDY}

\subsection{Derivation}

We investigate here the  impact of the planetary mass to the wavelength dependent transit depth, $C_{atm}$. Here, the goal is to use simple approximations to extract analytically the influence of each atmospheric/planetary parameter to the transit depth. We present in this section the key steps but the detailed derivation can be found in  Appendix.
We follow the approach taken by \cite{Brown_2001, fortney_clouds, Lecavelier_des_Etangs_2008, deWit_mass, Heng_non_iso_2015}. For a clear-sky atmosphere we have:

\begin{equation}
    C_{atm}(\lambda) = 2 \pi \int_{0}^{z_{max}} (R_0 + z) \left(1-\exp[\tau(z,\lambda)]\right) \text{d}z,
\end{equation}
where $R_0$ is the  radius at which the atmosphere becomes  opaque at all wavelengths,  $z$ the altitude from $R_0$ and $\lambda$ is the wavelength.
$\tau$ is the optical depth, i.e.:

\begin{equation}
    \tau(z,\lambda) = 2 \int_0^{x_f}\sum_i n_{0i}e^{-\frac{z}{H}}e^{-\frac{x^2}{2(R_0+z)H}}\sigma_i(p,T,\lambda) \text{d}x,
\end{equation}

where $x$ is the distance from the planet normal, $n_{0i}$ is number density of species $i$ at $z = 0$, $H$ is the scale height,  $\sigma_i$ is the cross section of the species $i$, $p$ is the pressure and $T$ the temperature. We can estimate the temperature and pressure dependence of the cross sections $\sigma$ by assuming a linear interpolation from tables of known values of $\sigma$, as currently done in most retrieval models (\cite{Hill_xsecT} and \cite{Barton_xsecP}). 

\begin{equation}
    \sigma_i(T) = \sigma_i(T_1)+\frac{\sigma_i(T_2)-\sigma_i(T_1)}{T_2-T_1}(T-T_1),
\end{equation}
\begin{equation}
    \sigma_i(p) = \sigma_i(p_j)+\frac{\sigma_i(p_{j+1})-\sigma_i(p_j)}{p_{j+1}-p_j}(p-p_j),
\end{equation}
where $T_1$, $T_2$, $p_j$ and $p_{j+1}$ are temperatures and pressures known from cross section tables. Since the pressure differences across the $x$ axis are large (larger than the interpolation intervals), we sum over intervals $(x_j, x_{j+1})$ of known pressures ($p_j, p_{j+1}$). 
This approximation allows us to derive analytically  the path integral  along the line of sight.

\begin{equation}
    \tau(z, \lambda) = \sum_j \sum_i n_{0i} e^{-\frac{z}{H}}\sqrt{\pi (R_0+z) H} \left( \sqrt{2} I_j\sigma_i(p_j,T_1) +K^{T}_{ij} \sqrt{2} I_j(T-T_1) + (K^{p}_{ij} + K^{X}_{ij} (T-T_1))(p_0 I_j' e^{-\frac{z}{H}} - \sqrt{2} I_j p_j) \right).
    \label{main_eq}
\end{equation}

The coefficients $K_{ij}^{T,p,X}$  are  the derivatives of the cross section with respect to either $T, p$ or both (their expression is given in the Appendix). $I_{j}$ represents the integration  of the opacity along the $x$ axis. 
Finally, the scale height $H$ is defined by:

\begin{equation}
    H = \frac{k_b T (R_0 + z)^2}{\mu M_p G},
\end{equation}

where $\mu$ is the mean molecular mass of the atmosphere, $G$ is the gravitational constant, $k_b$ the Boltzmann constant and $M_p$ the planetary mass.

\subsection{Interpretation}

The equations derived in the previous section can be used to predict the degeneracies we expect in retrieval simulations. Similar equations and degeneracies have been studied in previous works \citep{Brown_2001, deWit_mass, Griffith_degen_2014, fortney_clouds, Rocchetto_biais_JWST, Line_cloud_2016, Batalha_mass, Heng_theory_2017, Lecavelier_des_Etangs_2008, Tinetti_ariel, Fisher_38spec_2018, Welbanks_degen_2019}. We summarise here  the key findings which are relevant for our discussion on planetary mass. In general, the mass is expected to be well retrieved as its contribution to the transit depth calculation is  uniquely constrained by the atmospheric scale height. The mass appears only in the scale height definition, while the other parameters are constrained from other individual contributions to the opacity. 

\begin{itemize}
\item $R_0$ is the  radius at which a clear-sky atmosphere becomes  opaque at all wavelengths. In the case of a cloudy atmosphere (Grey clouds),  degeneracies may exist as  $R_0$ cannot be detected accurately below the cloud deck. 

\item For gaseous planets, $\mu$ is usually equal  to roughly $\sim2.3$, defined by the ratio H$_2$/He only. In secondary atmospheres, a wider range of main atmospheric components may exist and therefore $\mu$ is degenerate with $M_p$ in  equation \ref{main_eq}. 

\item The temperature has a similar role to the mass in the definition of the scale height (i.e: when an increase of the mass translates into a contraction of the atmosphere, a decrease of the temperature essentially plays the same role). However, the temperature is expected to change with altitude. Also, the temperature dependence of the cross sections could allow the temperature contribution to be distinguishable from the mass contribution if the observations are good enough and depending on the considered species and the atmospheric conditions.

\item The trace gases' number densities, $n_{0i}$ may change with altitude but otherwise are independent from the other parameters, including the mass. 
\end{itemize}

\section{RETRIEVAL ANALYSIS}

\subsection{Methodology}

In this section, we complement the analytical derivation in \S 2  with a number of relevant examples from retrieval simulations. We consider both primary and secondary atmospheres relevant to gaseous planets and super-Earths. We make extensive use of the open-source TauREx model (\cite{Waldmann_taurex1} and \cite{Waldmann_taurex2}) to simulate different atmospheric scenarios and perform  retrievals. TauREx is a fully Bayesian radiative transfer and retrieval framework which encompasses molecular line-lists from the ExoMol project (\cite{Tennyson_exomol}), HITEMP (\cite{rothman}) and HITRAN (\cite{gordon}). 

 For each case we begin by using TauREx in forward mode to generate a high-resolution theoretical spectrum. For the purpose of this investigation, we focus only on transit spectra and assume isothermal profiles. We will  cover eclipse spectra and more complex temperature-pressure profiles in a future work. Our model allows us to specify the main constituents of the atmosphere using their relative abundances (ratios of two molecules)

The high-resolution spectrum is convolved through the instrument model of \cite{mugnai_Arielrad} to simulate a spectrum as observed by ARIEL. Said synthetic spectrum acts as the input to the retrieval. In this study, we focus on observations obtainable with a  single transit, except in section 3.4 where we investigate the benefits on an increased signal to noise obtained by co-adding multiple transits. For each case considered we perform two retrievals: in the first case, the planetary mass is assumed to be known; in the second, it is retrieved as a free parameter. The latter  allows us to investigate whether the mass can be reliably estimated from transit spectra and assess the impact of mass uncertainties onto the retrieval of other atmospheric properties, such as the concentration of the trace gases, the temperature and cloud pressure.

In section 3.2, we investigate the case of a hypothetical hot-Jupiter, with parameters based on HD\,209458\,b (see table 1 in Appendix). We first present the case of a clear atmosphere and then extend the study to consider the impact of clouds at different cloud pressures.
In section 3.3, we investigate the case of a hypothetical super-Earth with a heavy atmosphere containing a significant fraction of $N_2$ or any other inert gas which cannot be detected through the identification of spectroscopic features. 
Section 3.4 is dedicated to the impact of the Signal to Noise on the retrieved mass, while in section 3.5 we compare the results of mass retrievals on HST data. 
Finally, we consider key examples of secondary atmospheres with clouds.

\subsection{Retrievals of gaseous planets}

The first set of retrievals focus on primary atmospheres, i.e. composed mainly of $H_2$, $He$. The simulated hot-Jupiter  is based on HD\,209458\,b and its parent star: the stellar and planetary parameters have been taken from \citep{Stassun_planetparam}. For trace gases, we have included   $H_2O$, $CH_4$ and $CO$, with mixing ratios  $10^{-5}$, $5 \times 10^{-6}$ and $10^{-4}$ respectively (e.g. \cite{Tsiaras_pop_study_trans}). We first simulate a clear atmosphere case, and then investigate the behaviour of the retrievals when clouds are present by varying the pressure of the cloud deck. 

\begin{figure}[h]
\centering
    \includegraphics[align=c,width=0.5\textwidth]{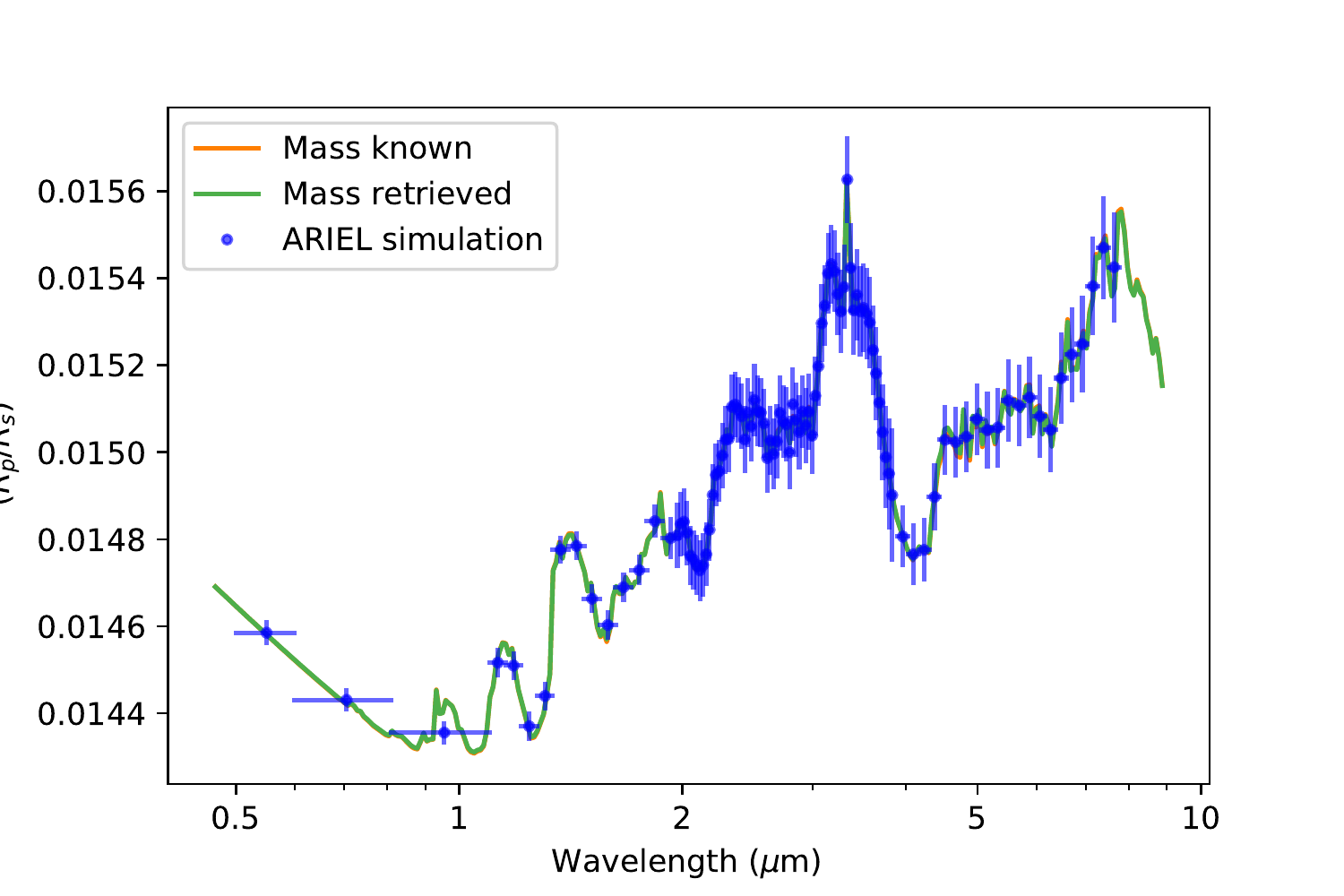}
    \includegraphics[align=c,width=0.49\textwidth]{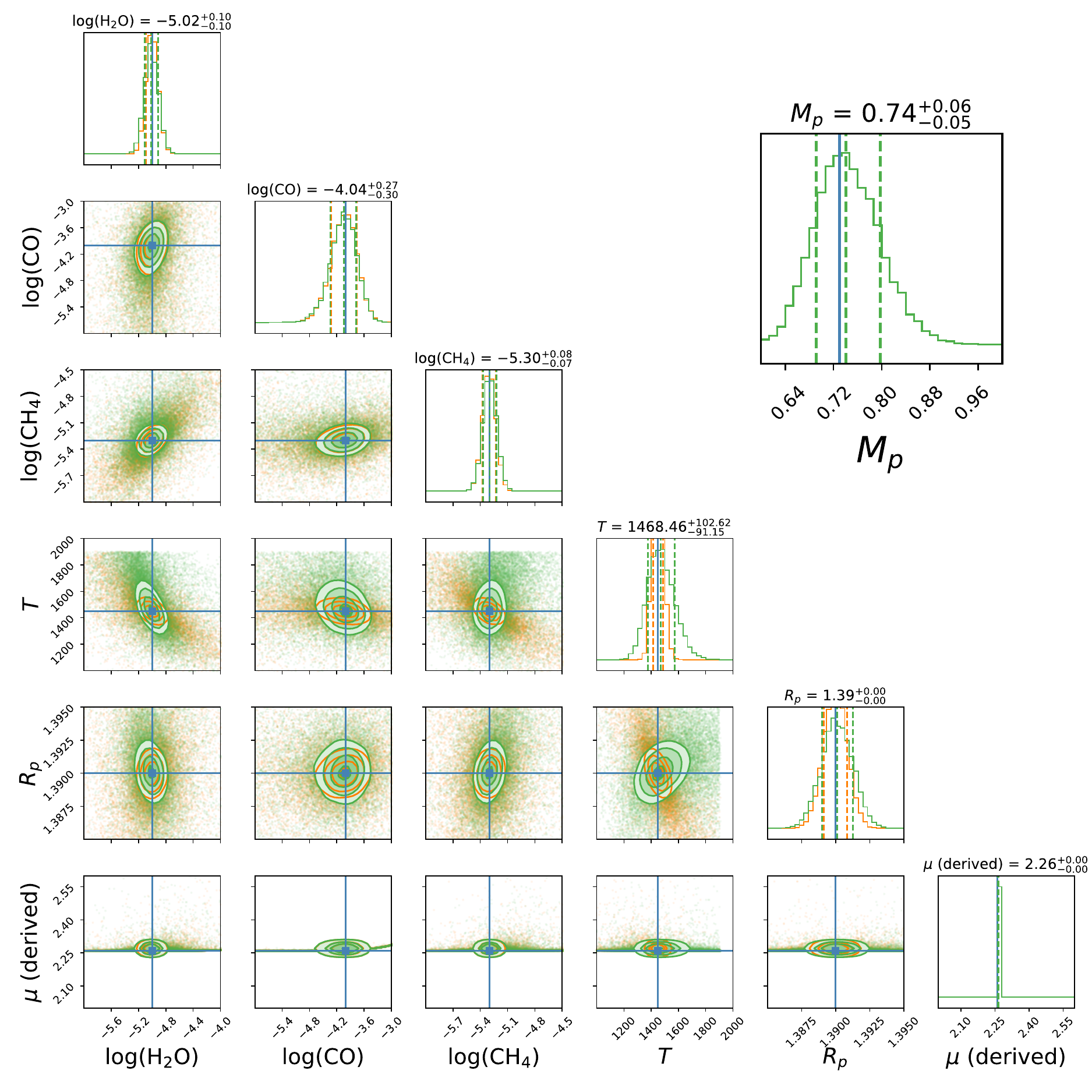}
    \caption{Spectra (left) and posteriors distribution (right) for a hot-Jupiter with a clear-sky atmosphere. Orange plots: the mass is known. Green plots: the mass is retrieved. The blue crosses indicate either the simulated ARIEL observations (left plot) or the ground truth values (right plot).}
    \label{fig:post_clear}
\end{figure}

\paragraph{Clear atmosphere} The fitted spectra and posteriors for both retrievals (`mass known' and `mass retrieved') in the case of a clear atmosphere are shown in Figure \ref{fig:post_clear}. Here the predictions of our analytic derivation still hold: molecular abundances and other parameters exhibit the same posterior distributions, showing that in this case the knowledge of the mass does not impact the results. The 1-sigma mass uncertainty corresponds to about  7\% of its  value. This uncertainty is propagated to the temperature posteriors, which are slightly larger when the mass is retrieved. However, the temperature is still very well constrained.

\begin{figure}[h]
\centering
    \includegraphics[align=c,width=0.32\textwidth]{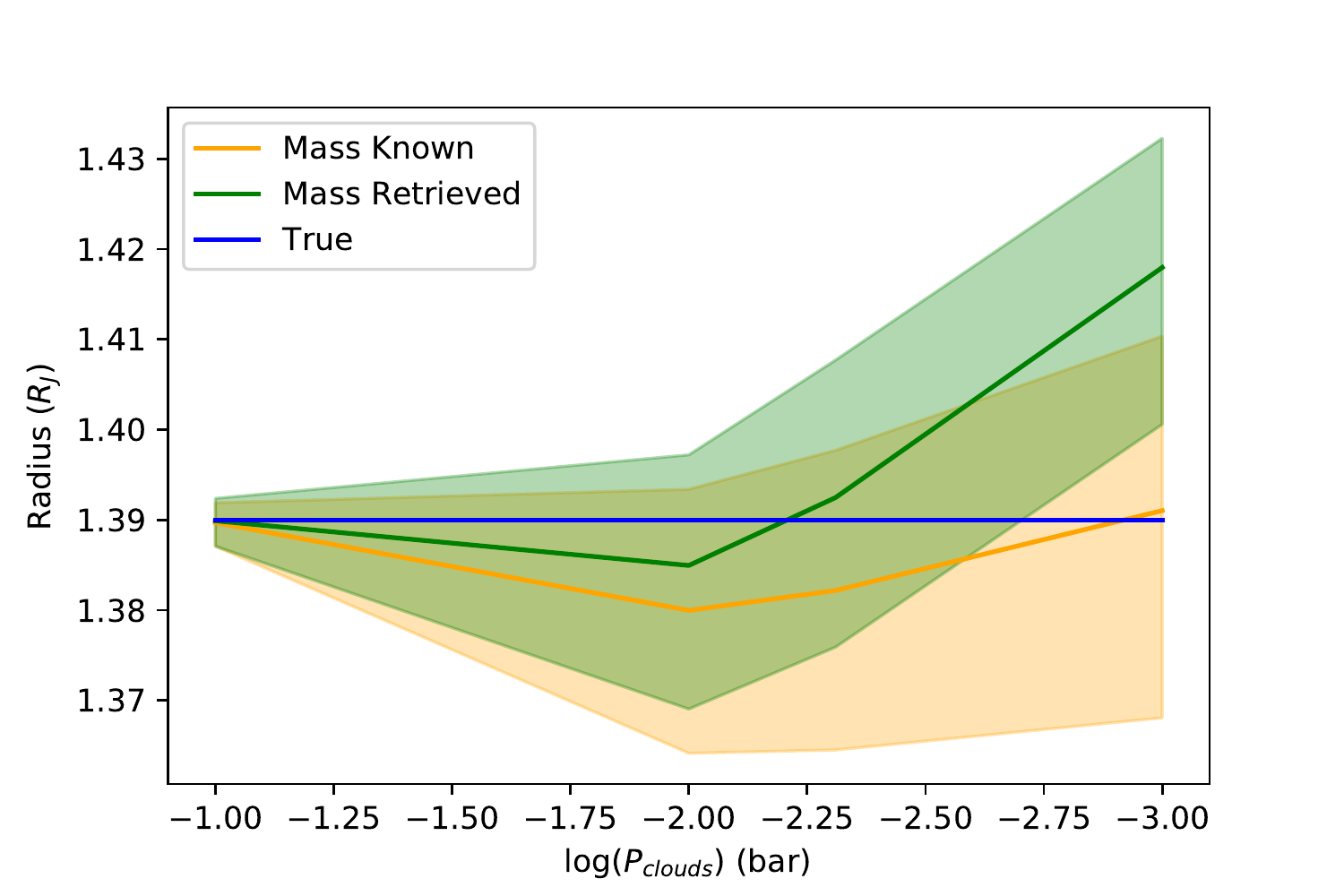}
    \includegraphics[align=c,width=0.32\textwidth]{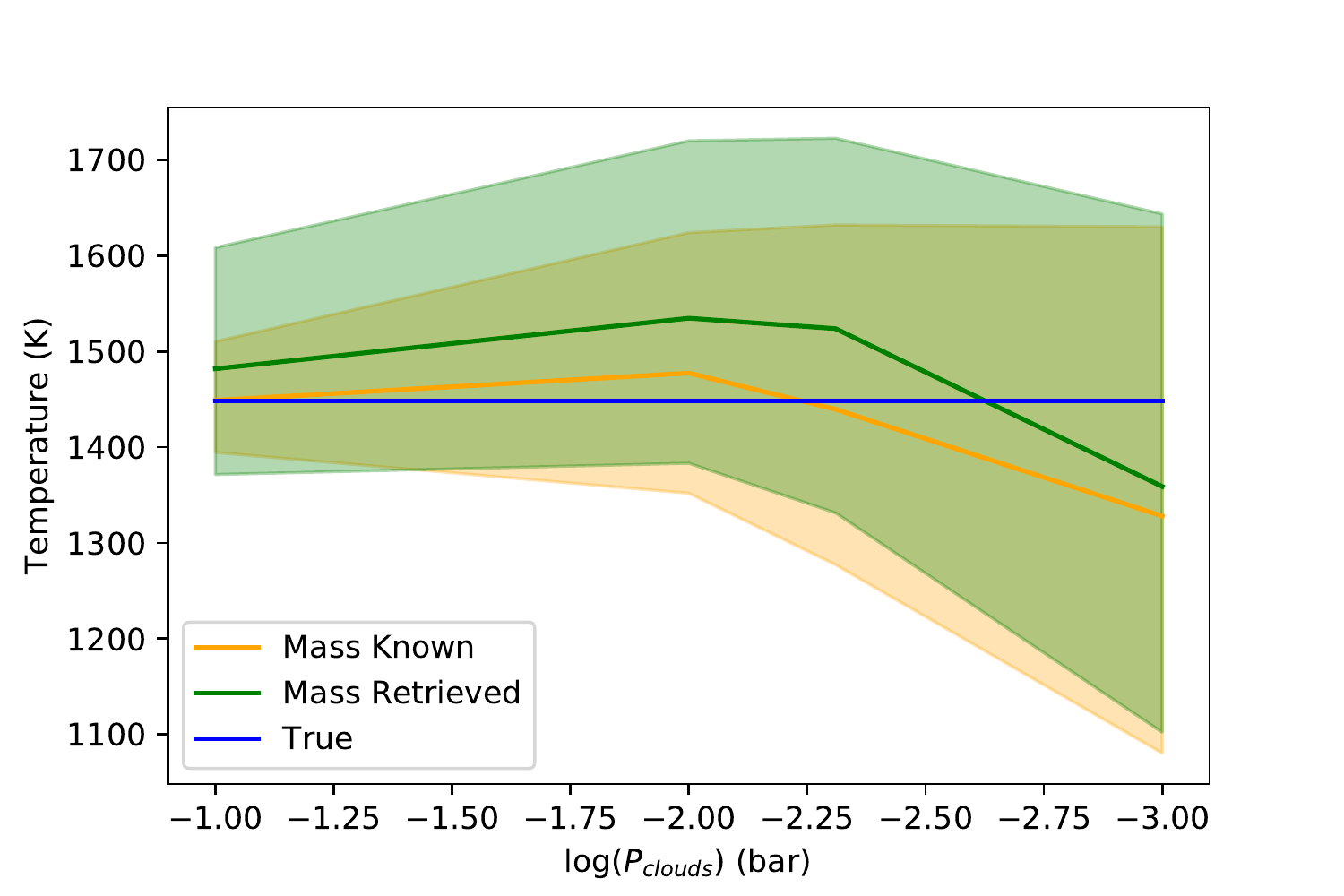}
    \includegraphics[align=c,width=0.32\textwidth]{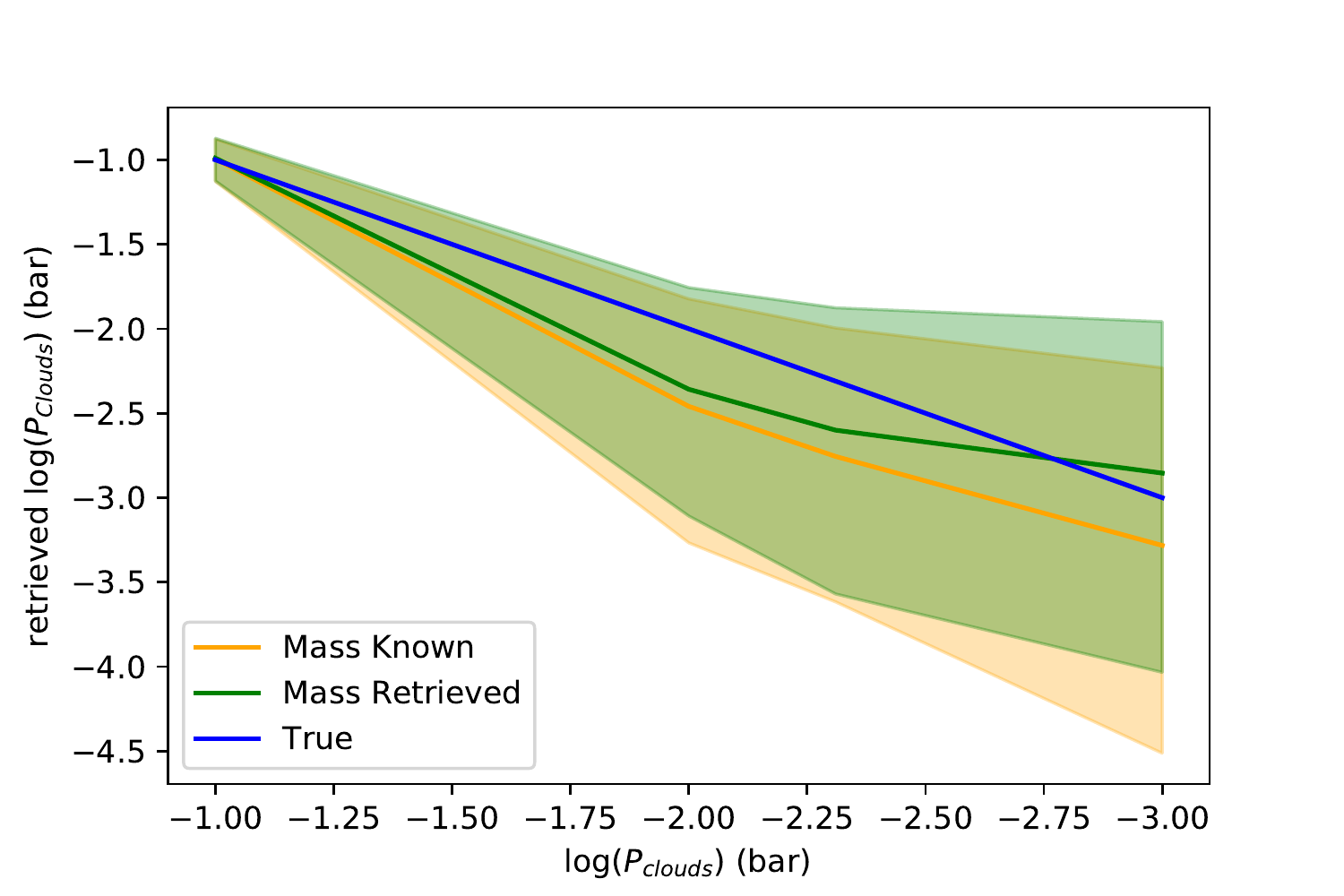}
    \includegraphics[align=c,width=0.32\textwidth]{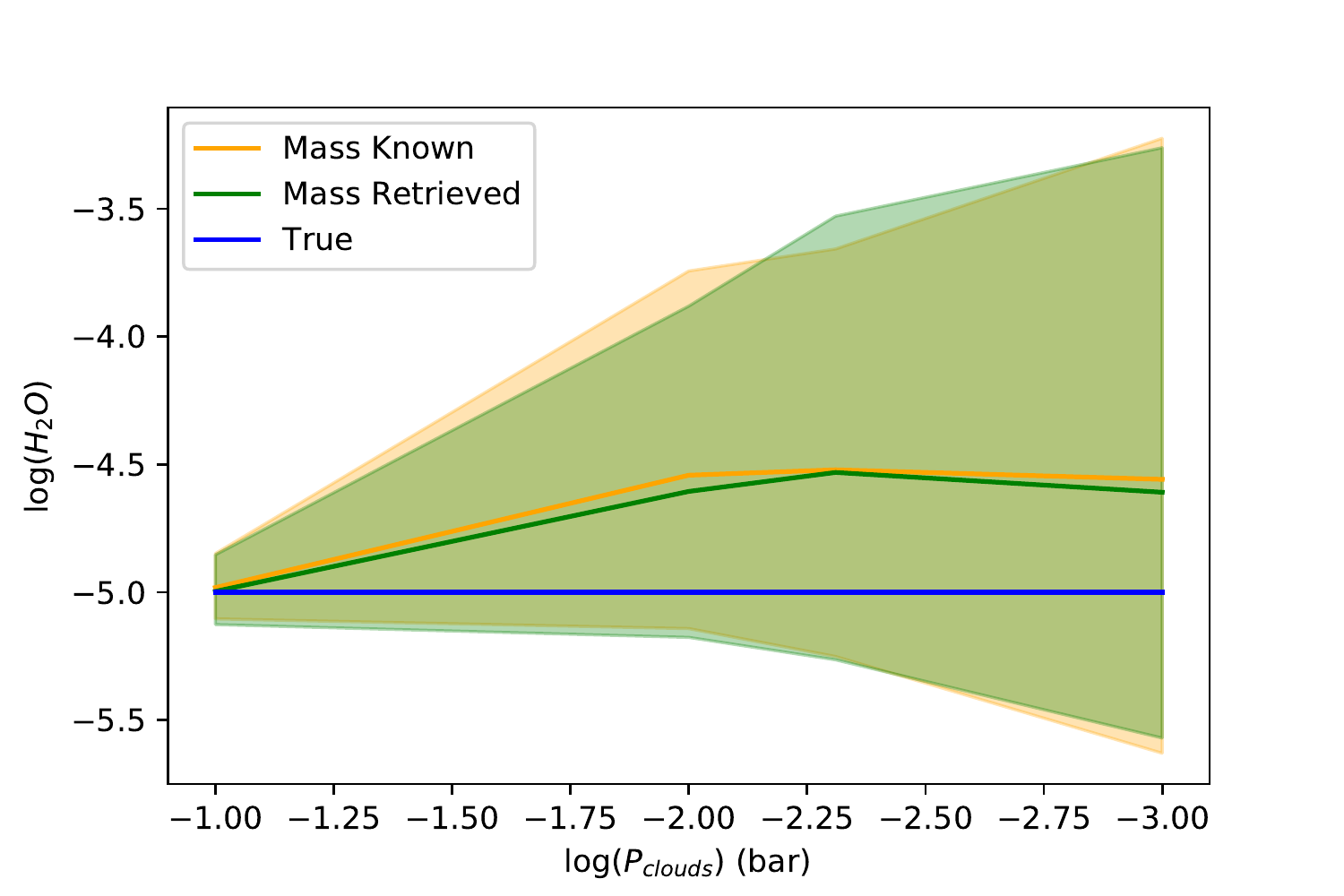}
    \includegraphics[align=c,width=0.32\textwidth]{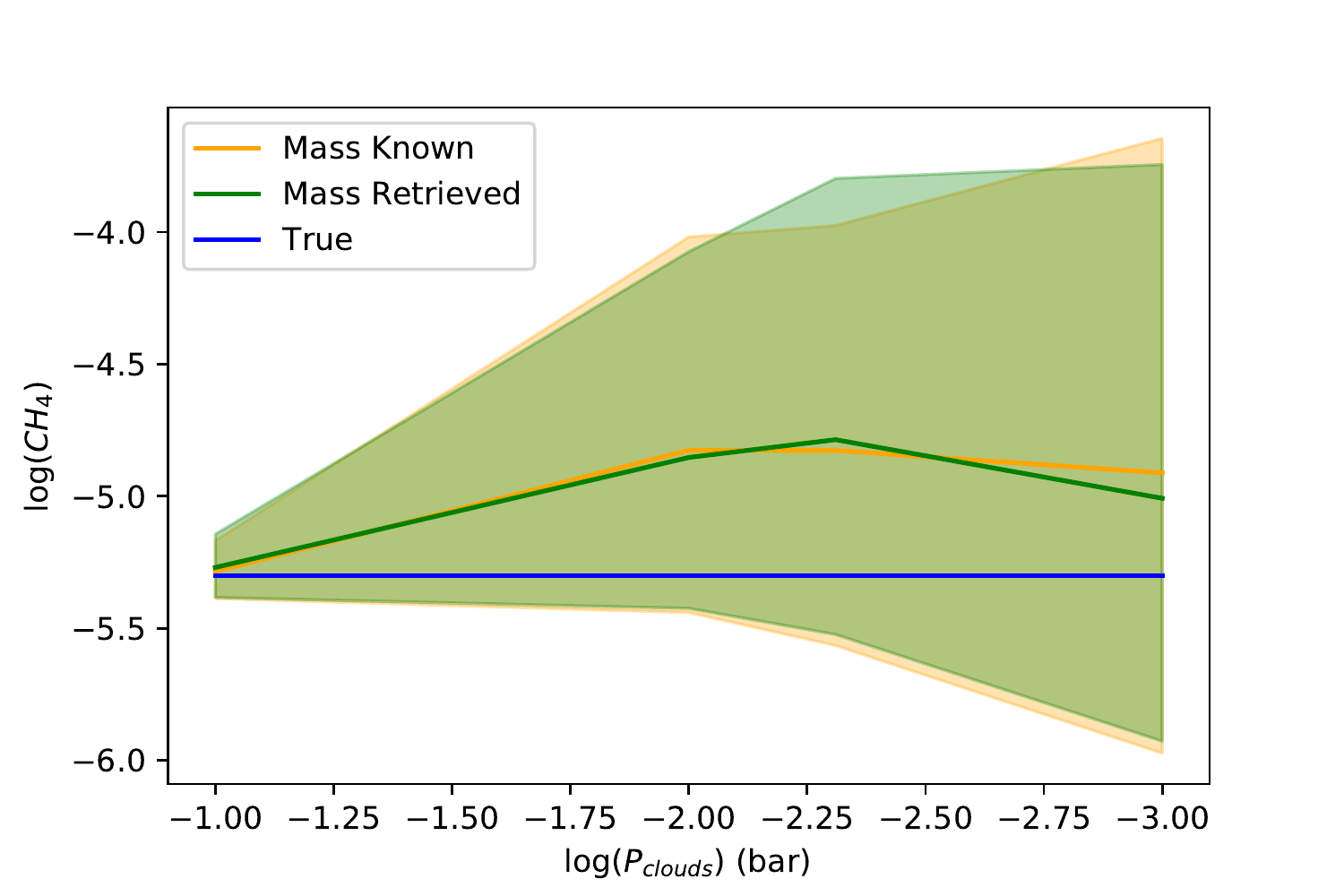}
    \includegraphics[align=c,width=0.32\textwidth]{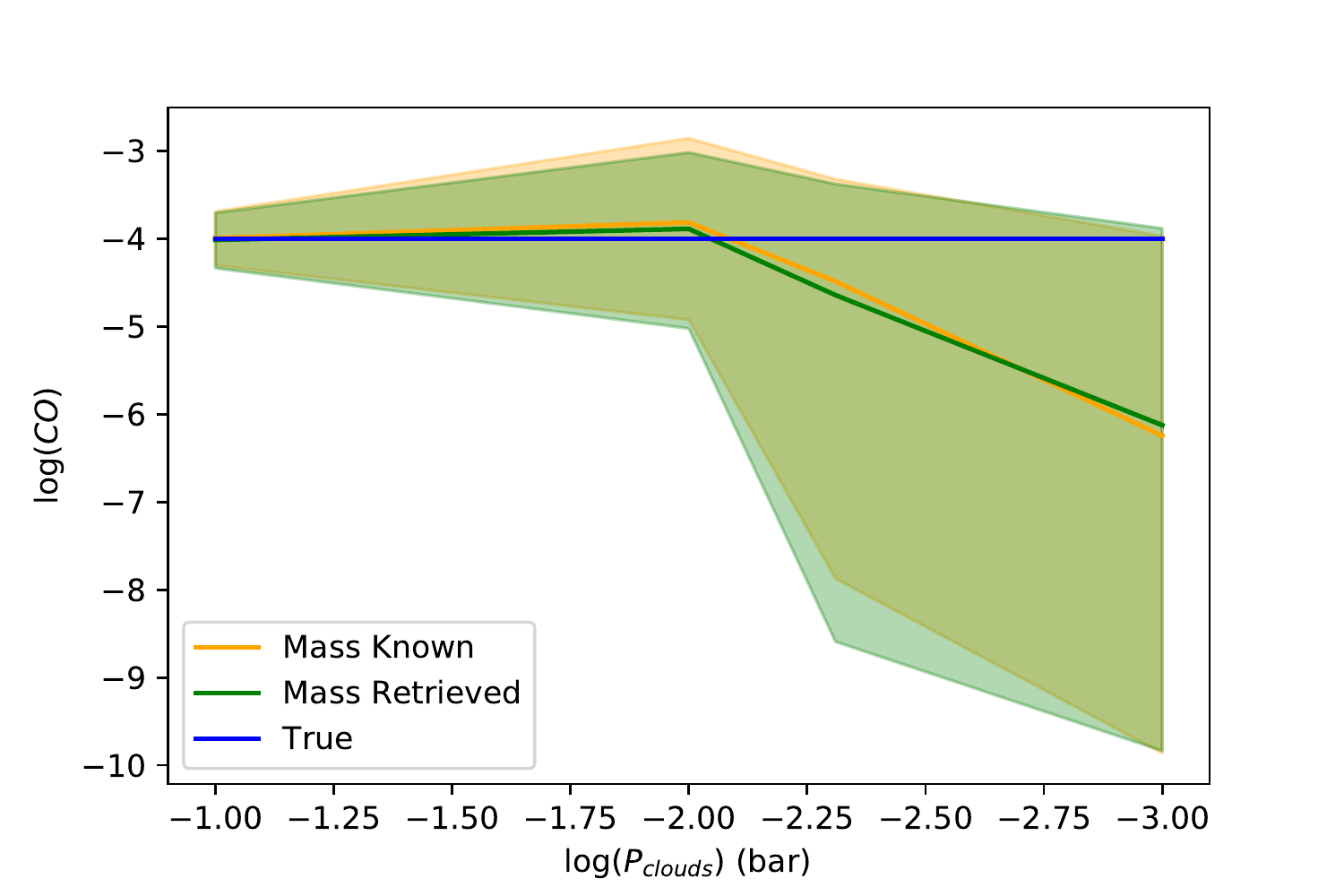}
    \caption{Comparison between the known/retrieved mass cases  as a function of cloud pressure. The  clear-sky case is rendered by placing the cloud deck at 10 bar.}
    \label{fig:retrieved_param_hj}
\end{figure}

\paragraph{Overcast atmosphere} Clouds are modelled by  including a completely opaque  cloud deck, where the cloud is optically thick below the cloud-top pressure. As mentioned previously, this choice represents the worst case scenario, due to  the maximum degeneracy with $R_0$, see equation \ref{eq:cloud} in the appendix. In addition to this issue, we note that being the observing time fixed and optimised for the clear sky case, for the high altitude clouds the signal to noise ratio  decrease noticeably. Five cases are considered in our analysis: 
\begin{enumerate}
  \item{Clear sky case}, see Figure \ref{fig:post_clear}.
  \item{Opaque cloud case at $10^{-1}$ bar}
  \item{Opaque cloud case at $10^{-2}$ bar}
  \item{Opaque cloud case at $5 \times 10^{-2}$ bar}
  \item{Opaque cloud case at $10^{-3}$ bar}
\end{enumerate}
In Figure \ref{fig:retrieved_param_hj} we plot the comparison between the known/retrieved mass cases  as a function of cloud pressure. Some discrepancies appear only in the retrieval of the radius when the cloud pressure gets closer to  $10^{-3}$ bar.
For all the other atmospheric parameters,
the knowledge of the mass does not impact the retrieved values nor the uncertainties. While  the uncertainty of the retrieved  values increases when the cloud pressure decreases, as expected,  we do not observe a difference between the known and retrieved mass cases. The retrieved trace-gas abundances and temperature are within 1-sigma of the true value.  

\begin{figure}[h]
\centering
    \includegraphics[align=c,width=0.7\textwidth]{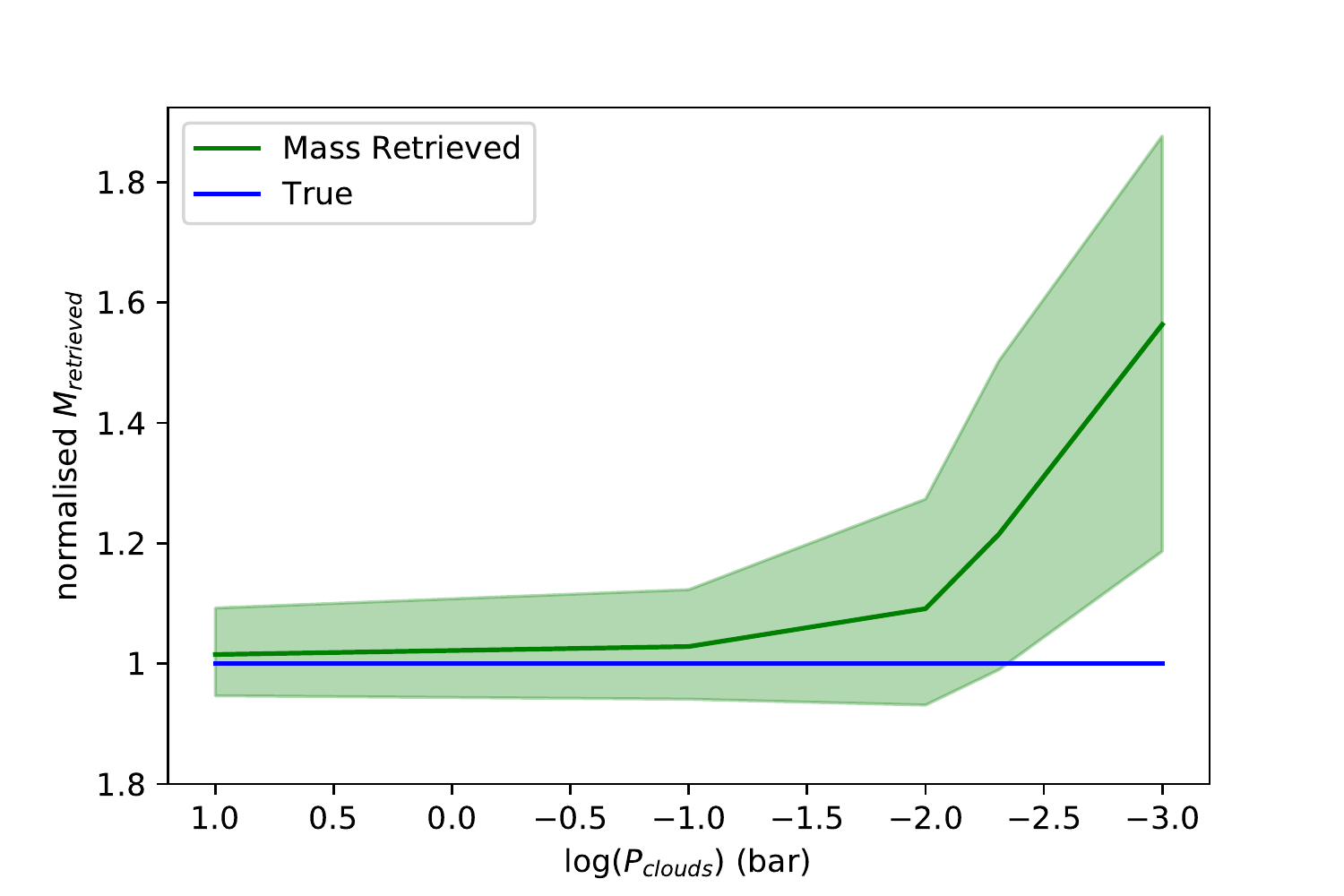}
    \caption{Normalised retrieved mass in the case of a gaseous planet as a function of cloud pressure. The green curve is the retrieved mass with its 1-sigma uncertainty. The blue line is the real value. The clear case is represented by a cloud deck at $10$ bar. The retrieved mass is not affected by low altitude clouds ($P_{clouds} \approx 0.1$ bar), while for high altitude completely opaque clouds, the retrieved mass starts to diverge from its true value (60 \% for $P_{clouds} = 10^{-3}$ bar).}
    \label{fig:Rmass_clouds}
\end{figure}

Focusing on the retrieval of the mass, the results of the normalised retrieved mass for each of the five cases are shown in Figure \ref{fig:Rmass_clouds}. We appreciate that the mass is well retrieved for all cases with clouds at low altitudes, while the  retrieved mass becomes less accurate when the cloud  pressure is lower than $10^{-2}$ bar. At the same time, the 1-sigma spread around the retrieved value also increases  with  the cloud altitude. 
The inaccuracy of the retrieved mass for high altitude, opaque clouds appears to be correlated with the inaccuracy of the retrieved radius, as shown in Figure \ref{fig:retrieved_param_hj}. To investigate further this important point, we discuss in detail a specific example of gas-giant planet with high altitude, opaque clouds. 

\begin{figure}[h]
\centering
    \includegraphics[align=c,width=0.5\textwidth]{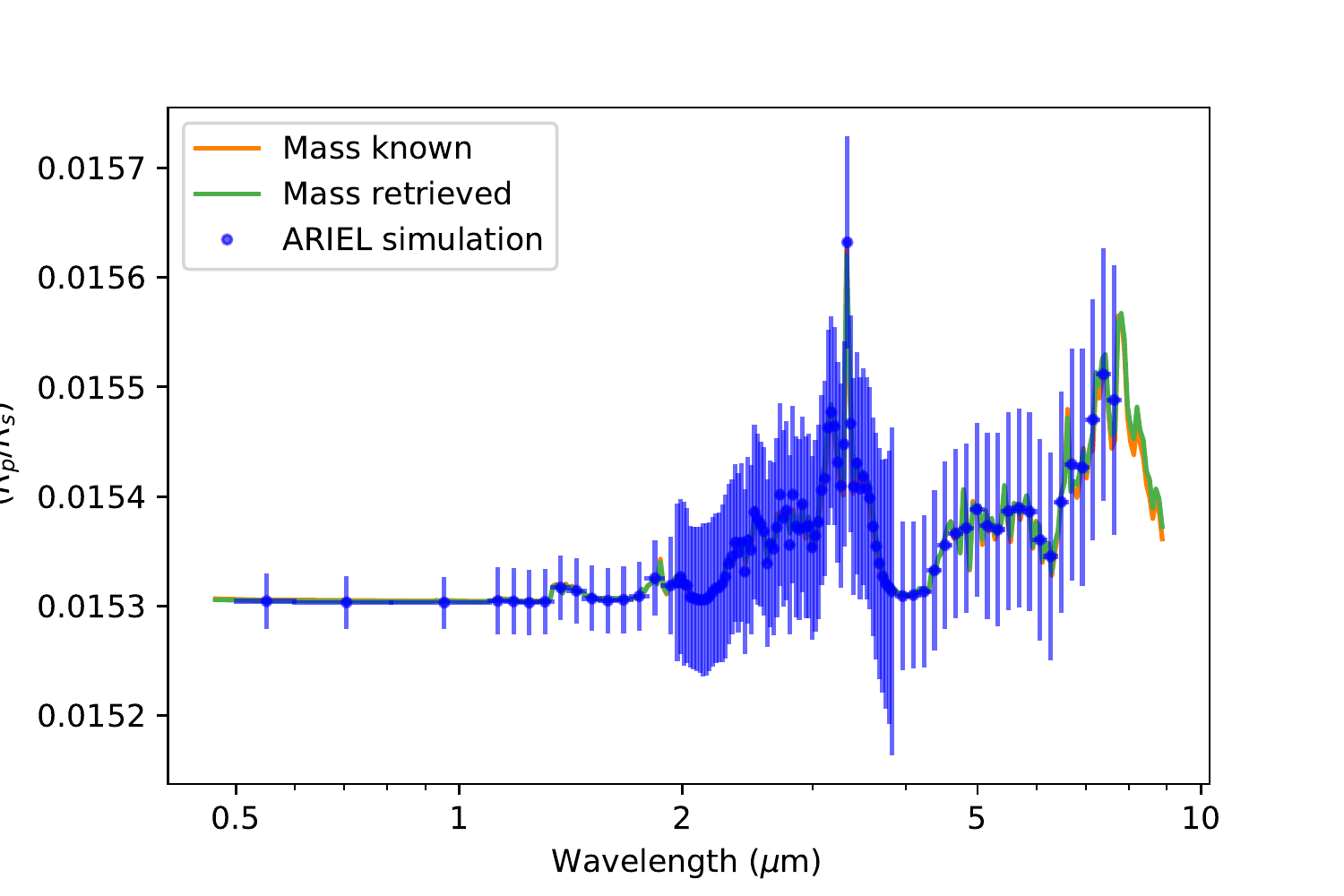}
    \includegraphics[align=c,width=0.49\textwidth]{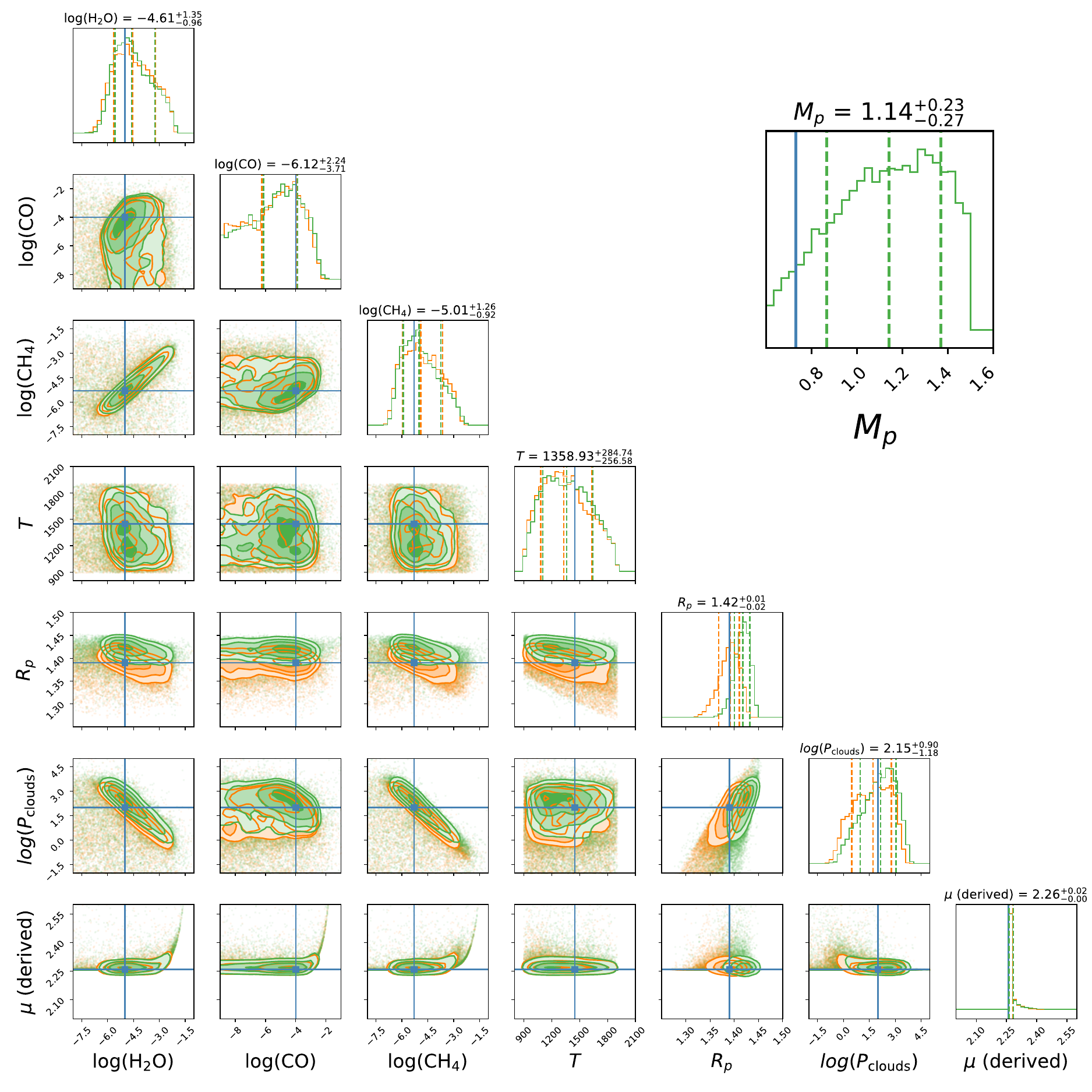}
    \caption{Spectra (left) and posteriors distribution (right) for a hot-Jupiter with  a cloudy atmosphere (opaque cloud deck at 10$^{-3}$ bar). Orange plots: the mass is known. Green plots: the mass is retrieved.  The blue crosses indicate either the simulated ARIEL observations (left plot) or the ground truth values (right plot).}
    \label{fig:post_clouds}
\end{figure}

\begin{figure}[h]
\centering
    \includegraphics[align=c,width=0.8\textwidth]{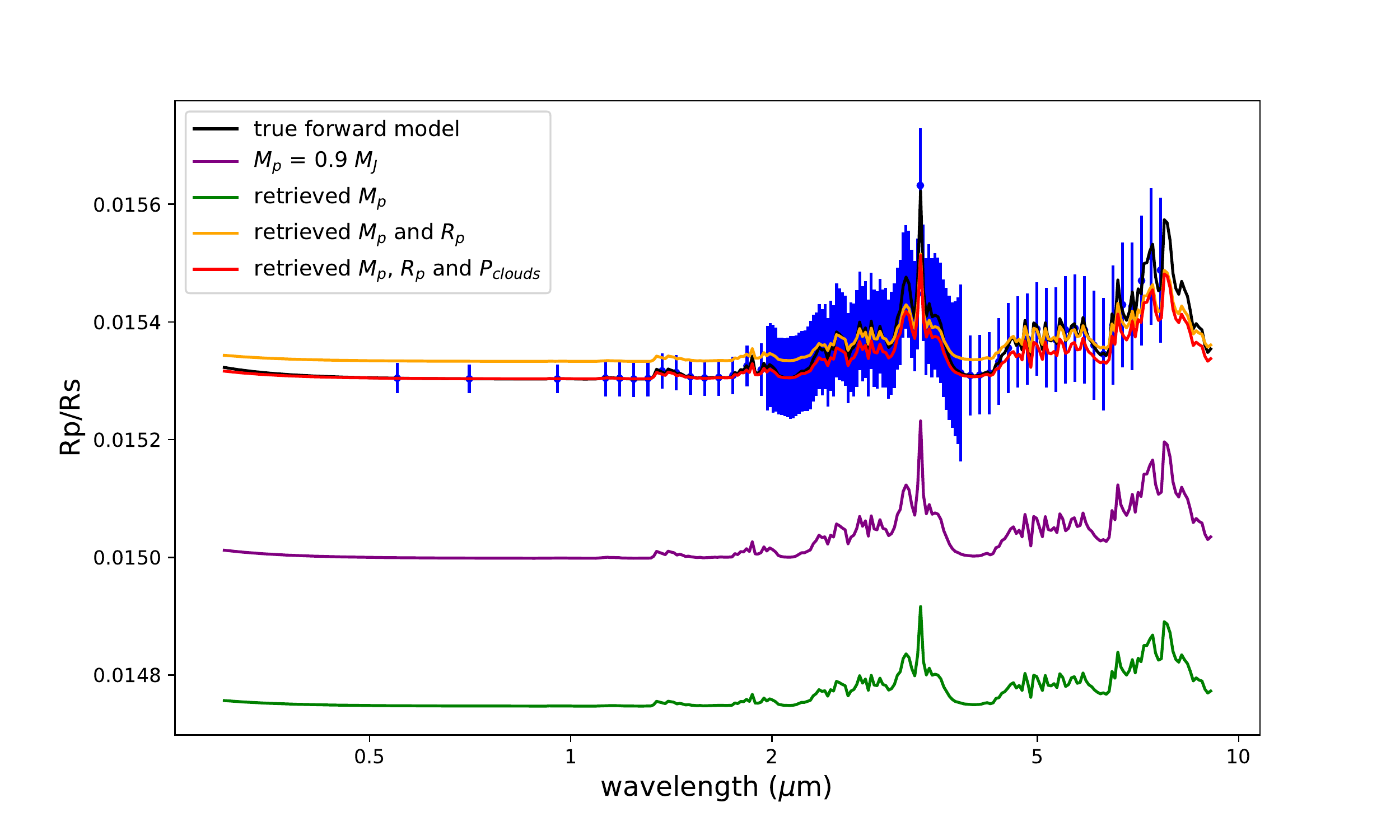}

    \caption{ Comparison of different forward models based on the cloudy case with cloud top pressure at 10$^{-3}$ bar. Black: True model. Purple: True model where only the mass is changed to $M_p = 0.9 M_J$. Green: True model where the mass is changed to the retrieved mean value. Orange: True model where the mass and the radius are changed to the retrieved value. Red: True model where the mass, the radius and the cloud pressure are changed to the retrieved value}
    \label{fig:forward_clouds}
\end{figure}

 Figure \ref{fig:post_clouds} illustrates an example where the cloud deck is located at $10^{-3}$ bar:  the `known' and `retrieved' mass scenarios are compared. As expected from Figure \ref{fig:retrieved_param_hj}, in this example the retrieved radius $R_0$ is no longer accurate and  the mass is no longer centred around its true value  and has large uncertainties.  More specifically, the retrieval shows a bias in selecting a larger radius $R_0$ to fit the spectrum. To compensate this bias, the mass retrieved is centred around a larger value, i.e. 
  1.14 $M_J$, compared to the true value, which is 0.88 $M_J$.  We illustrate the retrieval degeneracy between planetary mass, planetary radius and cloud top pressure by showing forward models for different cases in Figure \ref{fig:forward_clouds}. We note that, only small variations in the radius --less than 3\% -- are necessary to compensate for large mass offsets
  of $\sim$ 60\%. To mitigate this issue, the target could be observed for longer time to increase the signal to noise ratio, therefore reducing the level of degeneracies among these three parameters (see discussion in \S 3.4). 
  Again as expected from Figure \ref{fig:retrieved_param_hj}, we do not see significant differences in the other retrieved atmospheric parameters when the mass is known/unknown, which is reassuring for a mission or observing campaign dedicated to probe the atmospheric composition/thermal structure.

\subsection{Retrieval on secondary atmosphere planets}

In this section, we consider secondary atmospheres consisting of elements heavier than H/He. The super-Earth simulated here is taken from the ARIEL Target list \citep{Edwards_targetList}. The parameters used in our model are reported in the Appendix. We use the inactive gas N$_2$ to increase the mean molecular weight $\mu$ of the atmosphere and simulate a host of heavy atmospheres around a rocky planet. In our example, the atmosphere contains H$_2$O and CH$_4$ as trace gases: their absolute abundances are fixed at respectively $10^{-4}$ and $6 \times 10^{-4}$. The rest of the atmosphere is filled with a combination of H$_2$, He and N$_2$. By varying the N$_2$/He ratio, we essentially control the value of the mean molecular weight.

We have deliberately selected H$_2$, He and N$_2$ in our simulations,  so that the retrievals will not be guided by any spectral features of these molecules. This choice represents the worst case scenario to assess the degeneracy between the mass and the mean molecular weight. Atmospheres dominated by  species such as H$_2$O/CO$_2$/etc, would have traceable molecular features and would therefore represent a more favourable scenario for the inverse models. In this section, we consider the four following cases: 
\begin{itemize}
  \itemsep0em 
  \item $\mu = 2.3$ ($N_2/He$ = 0)
  \item $\mu = 5.2$ ($N_2/He$ = 1)
  \item $\mu = 7.6$ ($N_2/He$ = 2)
  \item $\mu = 11.1$ ($N_2/He$ = 4)
\end{itemize}

\begin{figure}[H]
\centering
    \includegraphics[align=c,width=0.7\textwidth]{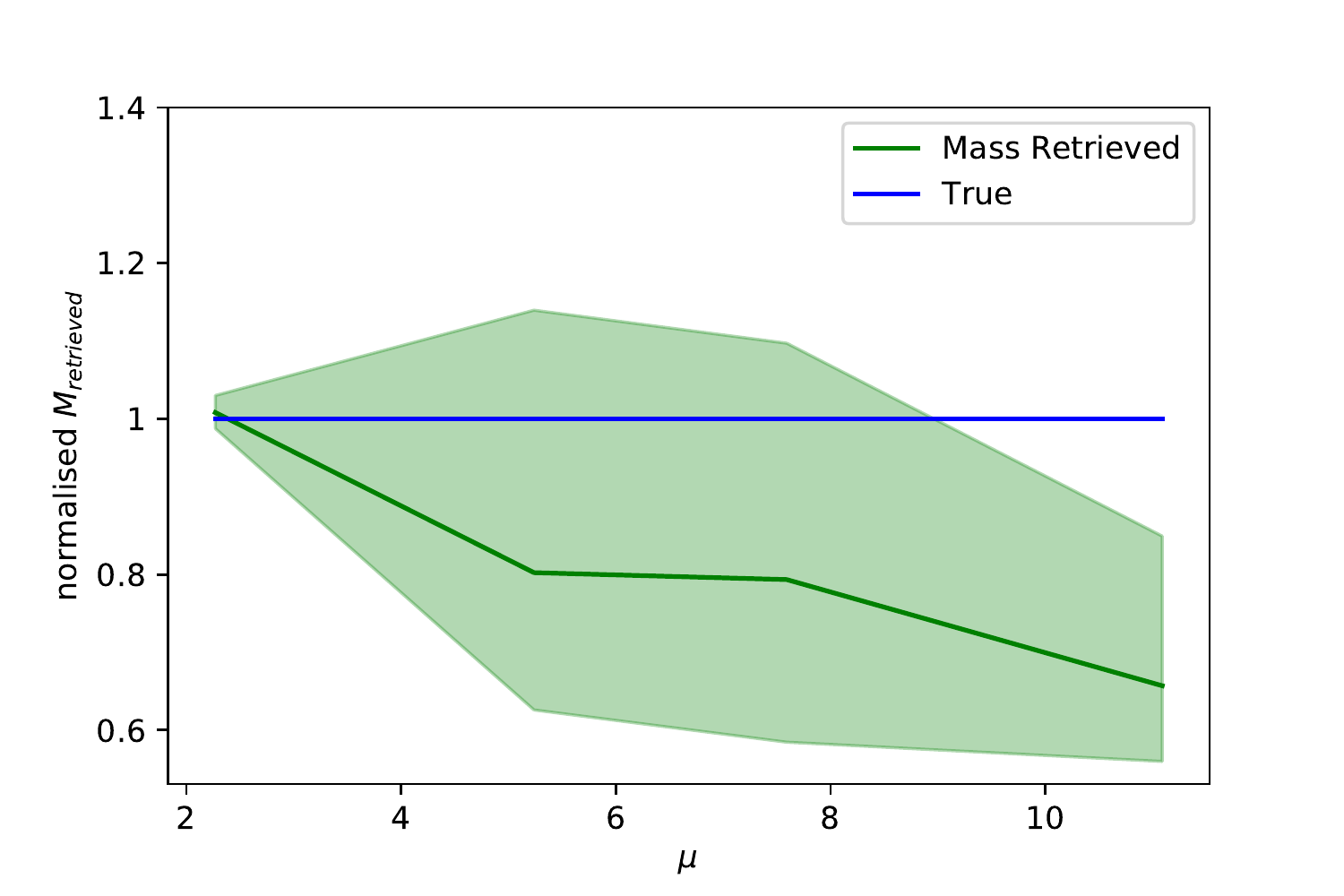}
    \caption{Normalised retrieved mass ($M_{retrieved}$ in green) for planets with a  secondary atmosphere  as a function of the mean molecular weight. The blue line represents the real value.}
    \label{fig:Rmass_heavy}
\end{figure}

We show in Figure \ref{fig:Rmass_heavy}  the normalised mass retrieved  as a function of the mean molecular weight $\mu$.

\begin{figure}[h]
\centering
    \includegraphics[align=c,width=0.5\textwidth]{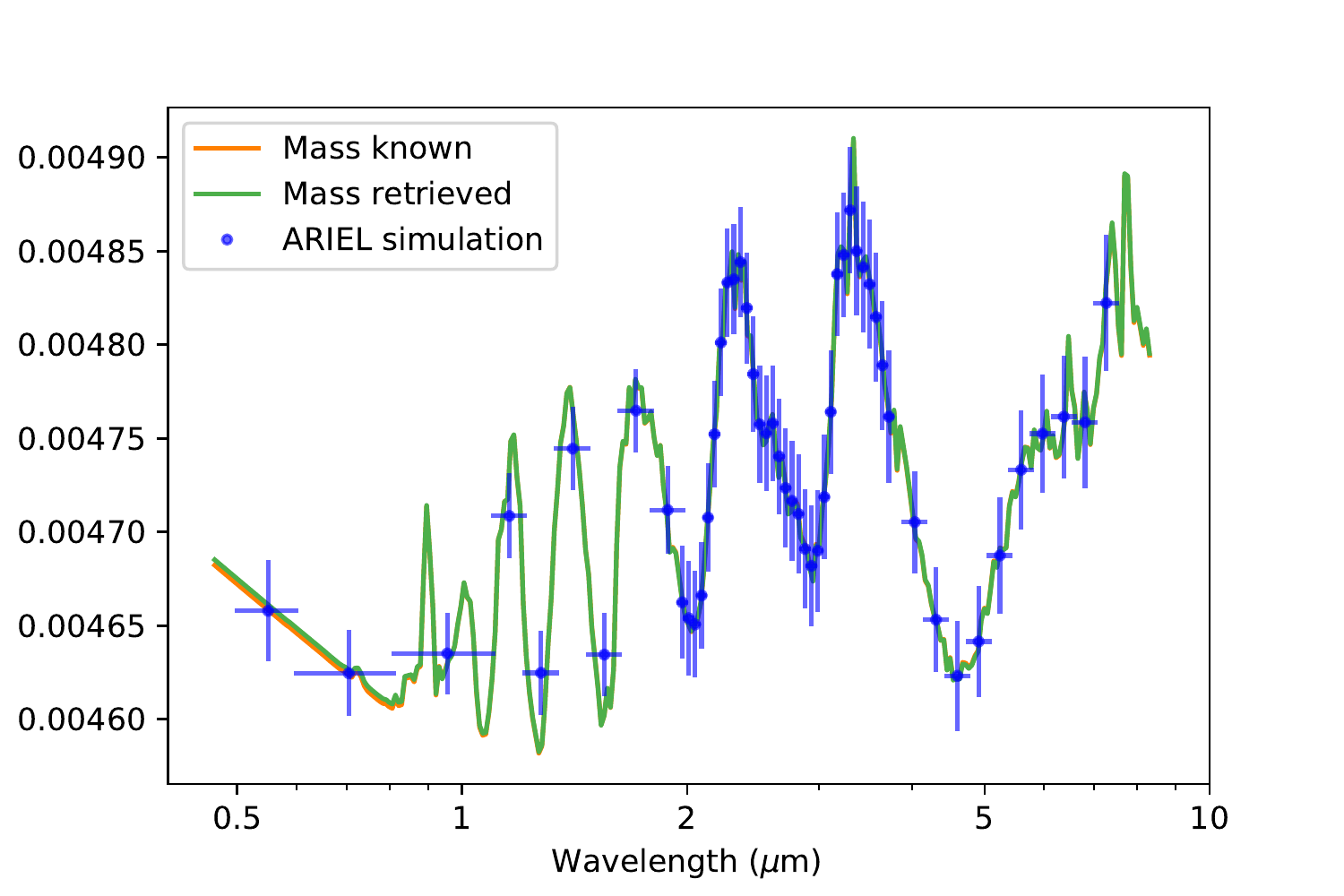}
    \includegraphics[align=c,width=0.49\textwidth]{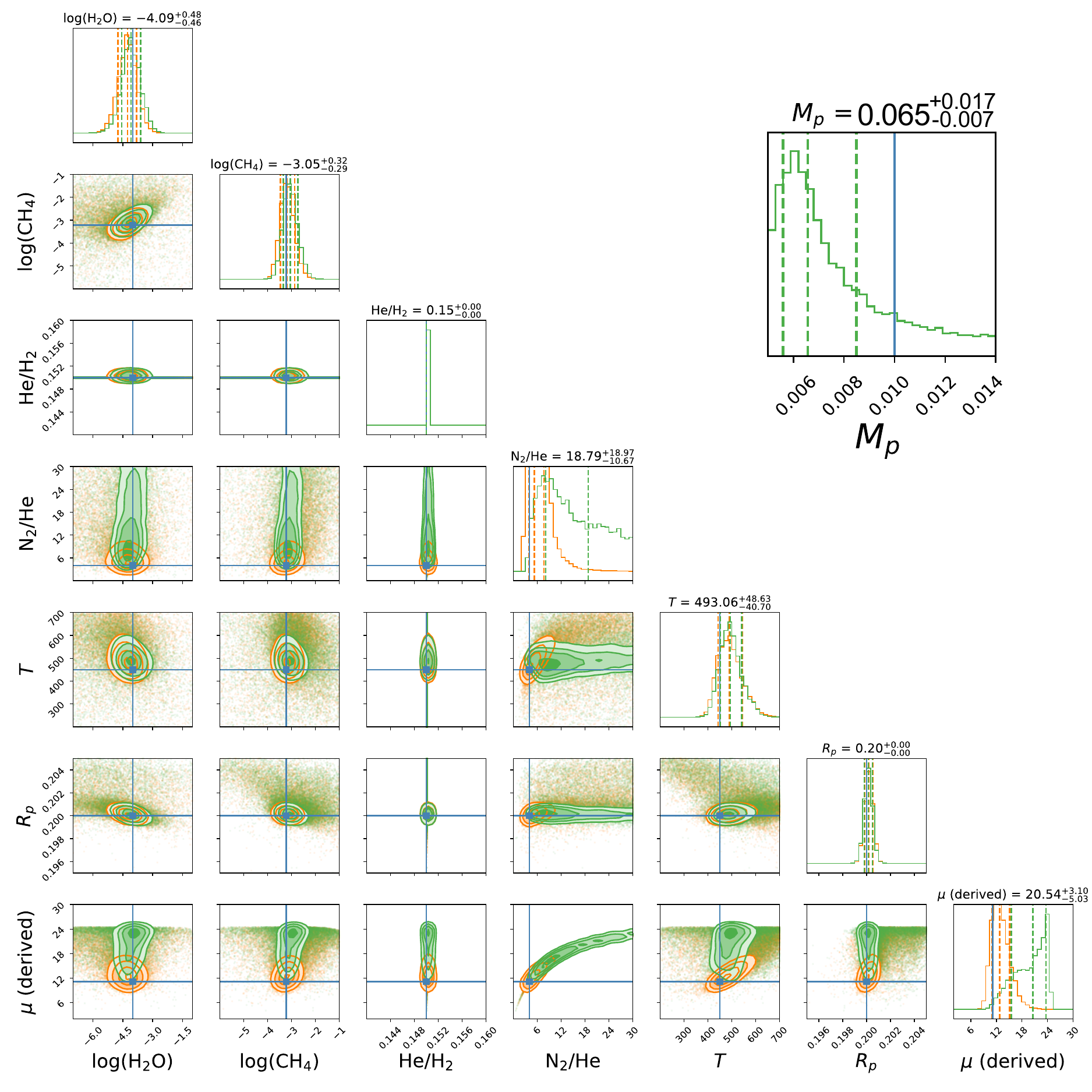}
    \caption{ARIEL simulated spectra (left) and posteriors distribution (right) for a cloud-free atmosphere with $\mu = 11.1$ (i.e. $N_2/He = 4$). Orange plots: the mass is known. Green plots: the mass is retrieved. Blue crosses:  simulated ARIEL observations obtained in one transit (left plot) and  true values in the posterior distributions (right plots).}
    \label{fig:spec_heavy}
\end{figure}

At small $\mu$, the atmosphere is dominated by a single gas species: $H_2$. This case has already been considered in \S 3.2. 
The degeneracy mass/mean molecular weight becomes  more important for increasing  $\mu$ (see Figure \ref{fig:Rmass_heavy}). For $\mu \ge 9$, the mass is not correctly retrieved due to the degeneracy predicted in \S 2. We show the case $\mu = 11.1$, i.e. $N_2/He = 4$, in Figure \ref{fig:spec_heavy}, which clearly illustrates the discrepancy between the ground-truth and the retrieved  $\mu$ and planetary mass. We find that the space of possible solutions for the retrieved mass in the case of secondary atmospheres is not centred around the true value. These results match the conclusions reached by \cite{Batalha_mass}: by analysing different cases of heavy atmospheres,  they found they could reproduce the same spectra with different sets of parameters. 
In our simulations, however, we show also that the trace gases, the temperature and the planetary radius are accurately retrieved with the same posteriors for both the  known and retrieved mass cases. 

\begin{figure}[h]
\centering
    \includegraphics[align=c,width=0.32\textwidth]{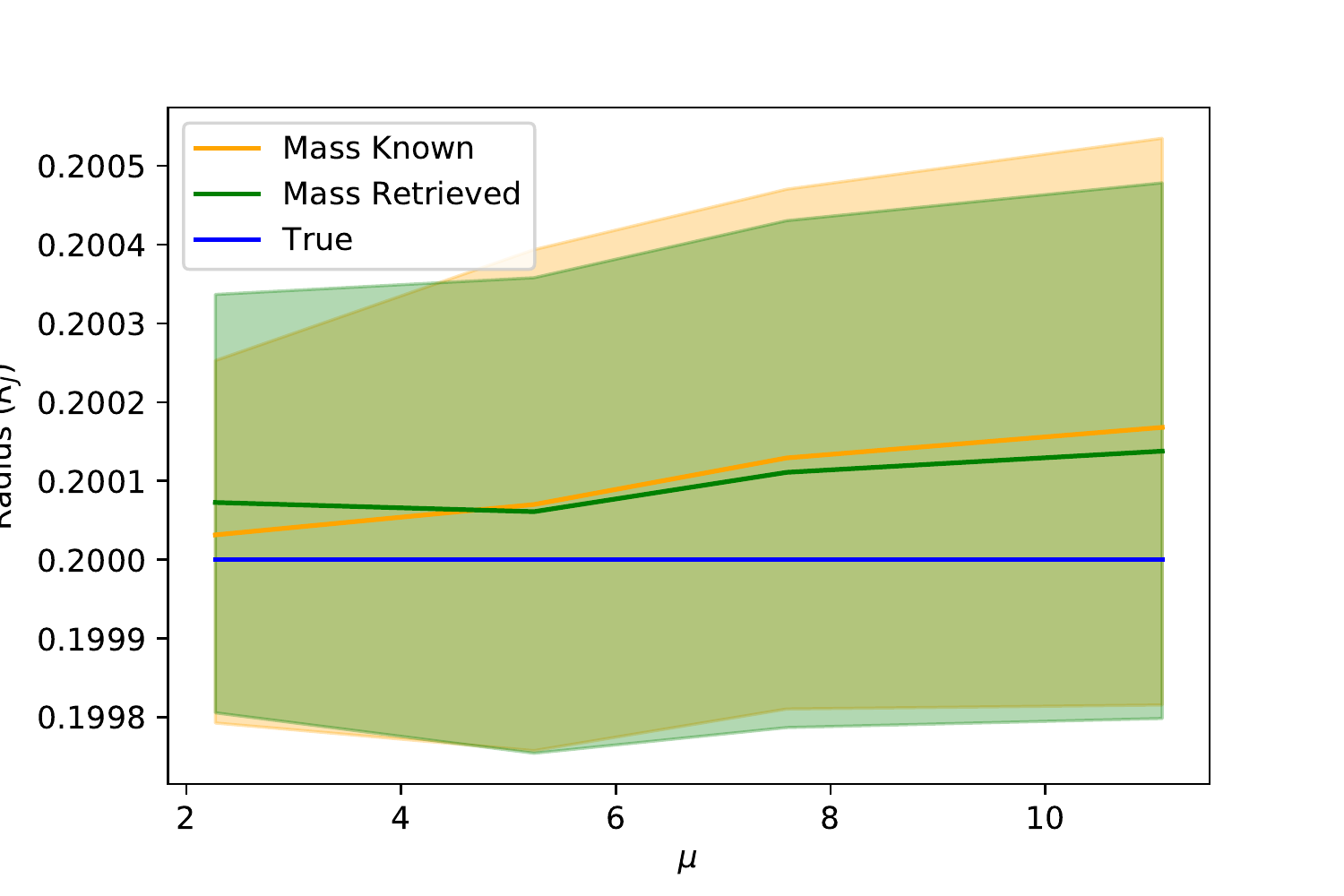}
    \includegraphics[align=c,width=0.32\textwidth]{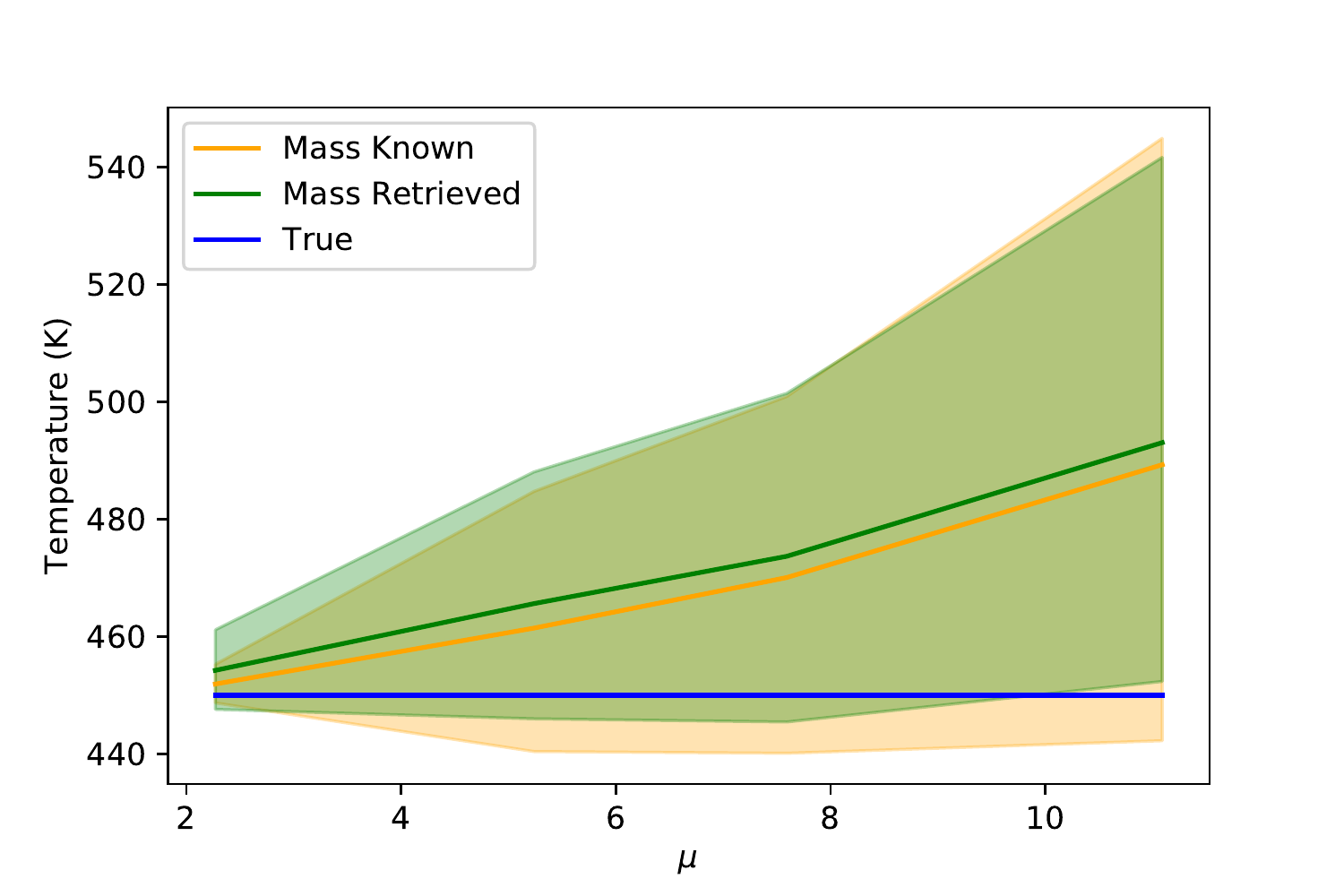}
    \includegraphics[align=c,width=0.32\textwidth]{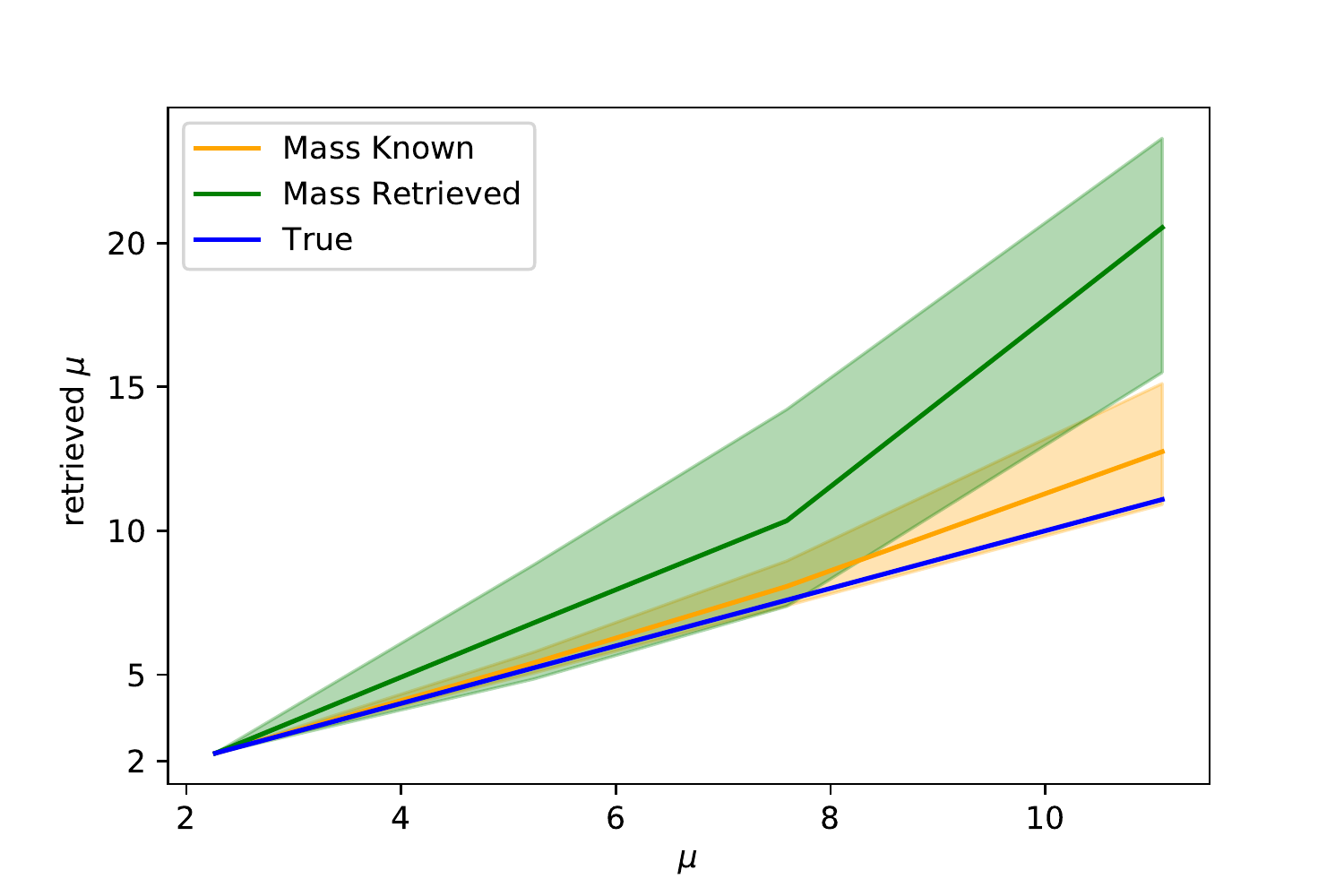}
    \includegraphics[align=c,width=0.32\textwidth]{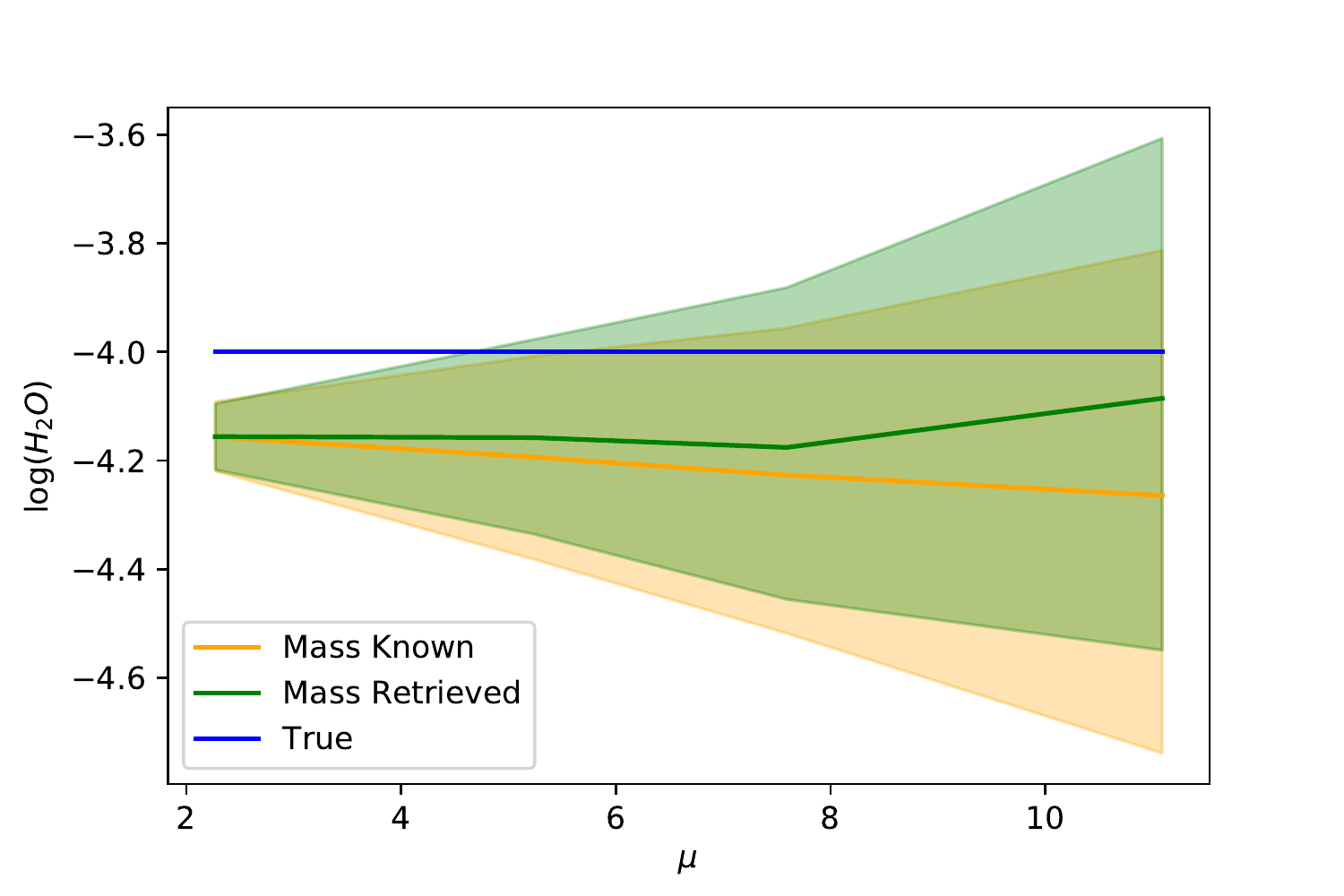}
    \includegraphics[align=c,width=0.32\textwidth]{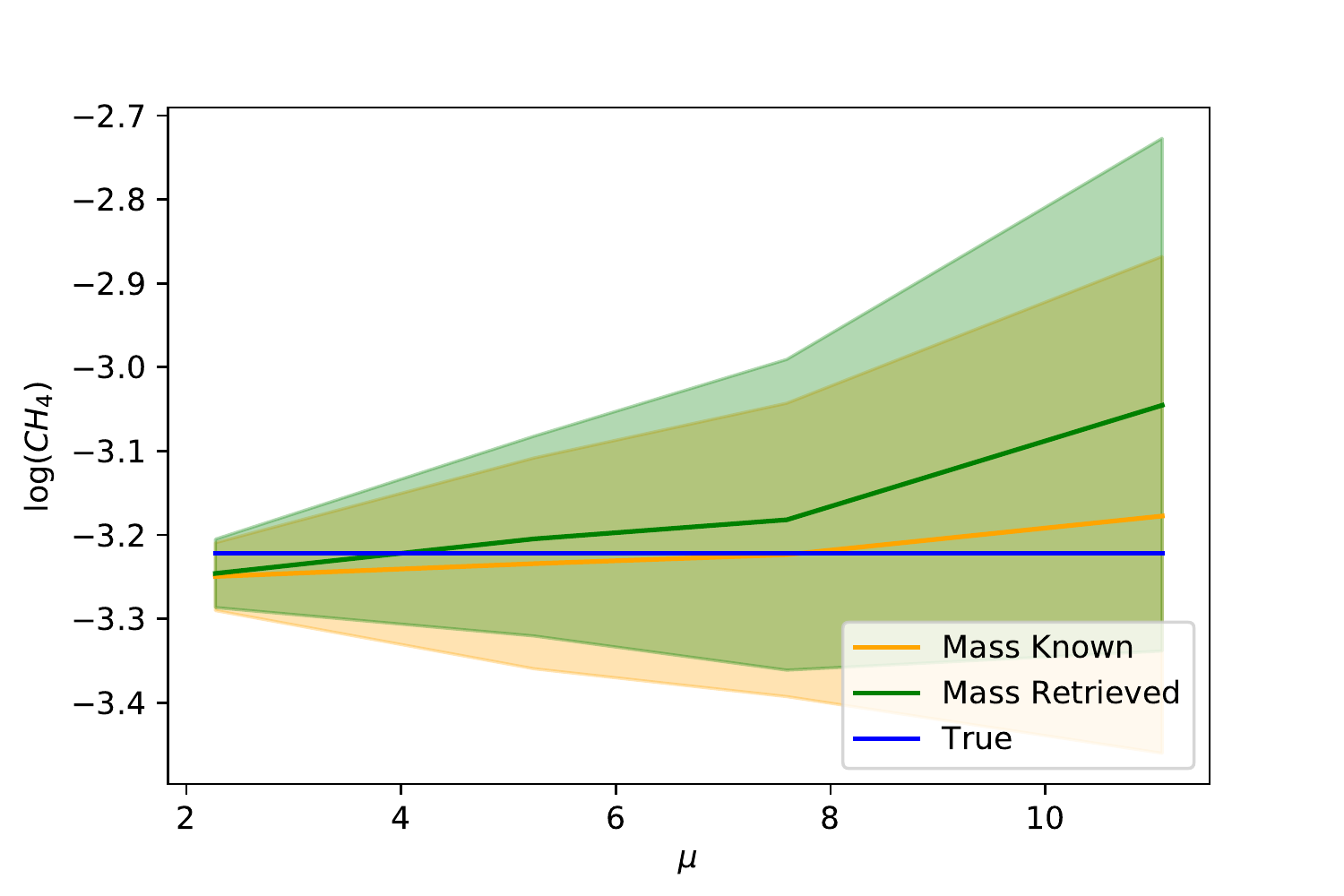}
    \caption{Impact of the mass  on the retrieval of the radius, temperature, mean molecular weight and trace-gas abundances for different scenarios of heavy atmospheres represented by increasing values of $\mu$. The simulated ARIEL observations are obtained in one transit}
    \label{fig:retrieved_params_se}
\end{figure}

We plot in Figure \ref{fig:retrieved_params_se},  a comparison of the retrieved parameters as a function of the mean molecular weight. This shows that the temperature, the trace gases and the radius have similar uncertainties when the mass is known and retrieved for  different values of $\mu$. The retrieved $\mu$ is degenerate with the mass and tends to be larger than the true value, hence the complementary smaller retrieved mass in Figure \ref{fig:Rmass_heavy}: the mean molecular weight and the mass are inversely correlated in these retrievals. Additionally, for all cases, the retrieved $\mu$ presents larger uncertainties when the mass is retrieved at the same time.

We conclude that for planets with a secondary atmosphere, an independent determination of the mass can help to break the degeneracy with the mean molecular weight.

\subsection{Importance  of the signal to noise and wavelength coverage of the transit spectrum to retrieve the  mass}

\begin{figure}[h]
\centering
    \includegraphics[align=c,width=0.7\textwidth]{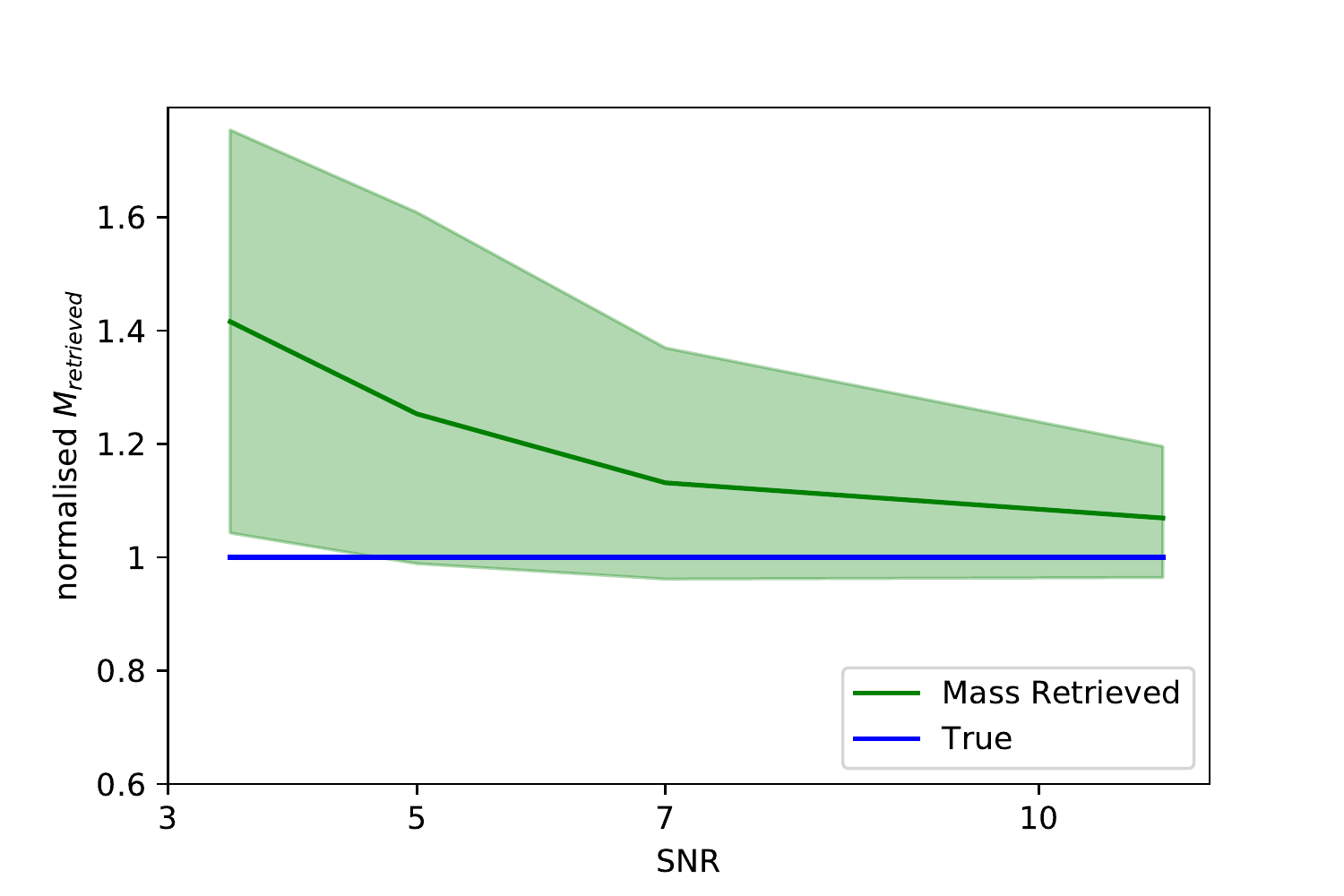}
    \caption{Normalised retrieved mass ($M_{retrieved}$ in Green) for a $N_2$-rich heavy atmosphere case ($\mu = 27.8$) as a function of S/N. Blue line: real value. }
    \label{fig:mass_stacks}
\end{figure}

The larger  $\mu$ is, the smaller is the spectral signal, and therefore it is important to guarantee an adequate S/N when we observe heavy atmospheres. We show
in Figure \ref{fig:mass_stacks} an example  of secondary atmosphere with large amount of N$_2$ ($\mu = 27.8$):  we plot the normalised retrieved mass error as function of the S/N.

\begin{figure}[h]
\centering
    \includegraphics[align=c,width=0.5\textwidth]{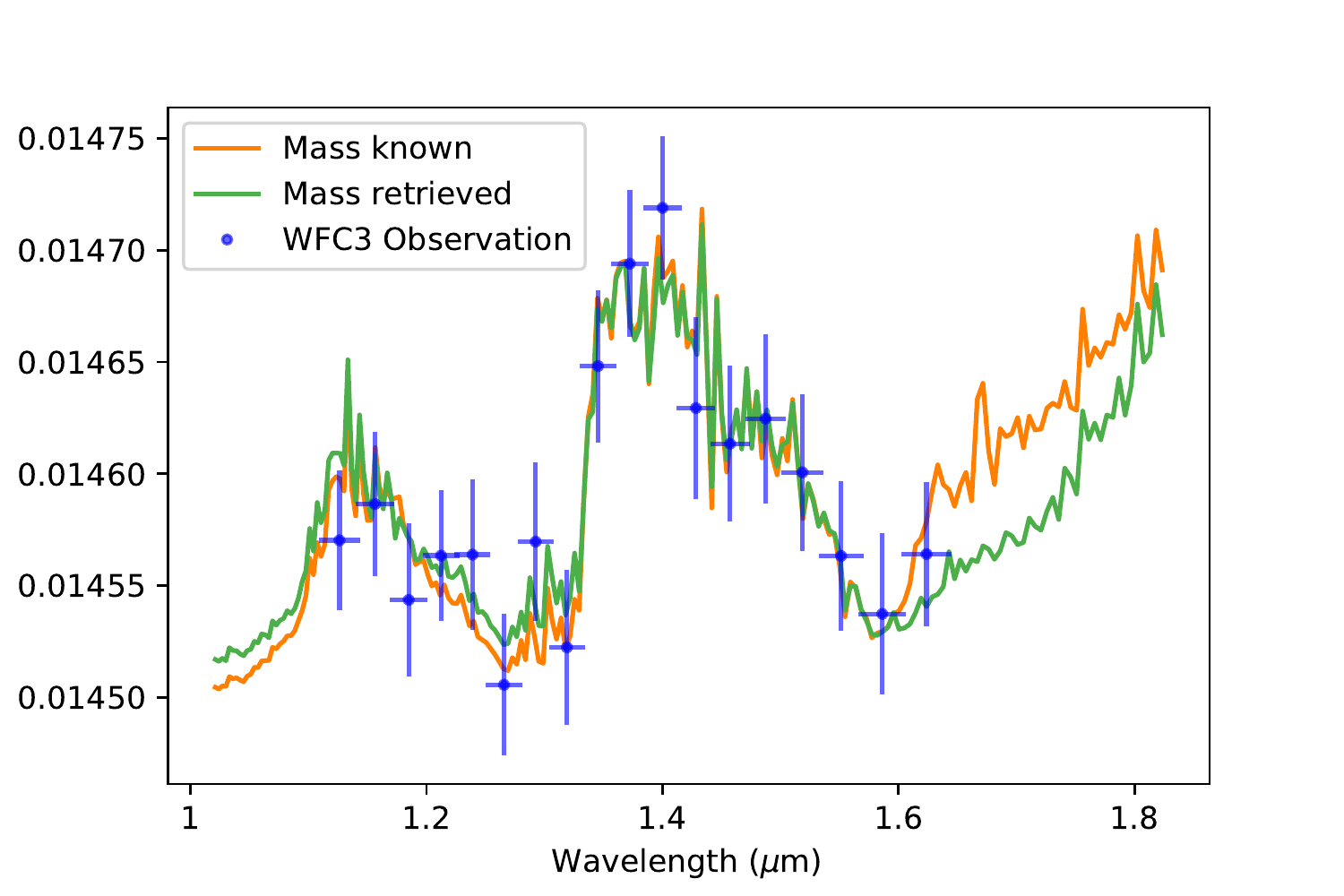}
    \includegraphics[align=c,width=0.49\textwidth]{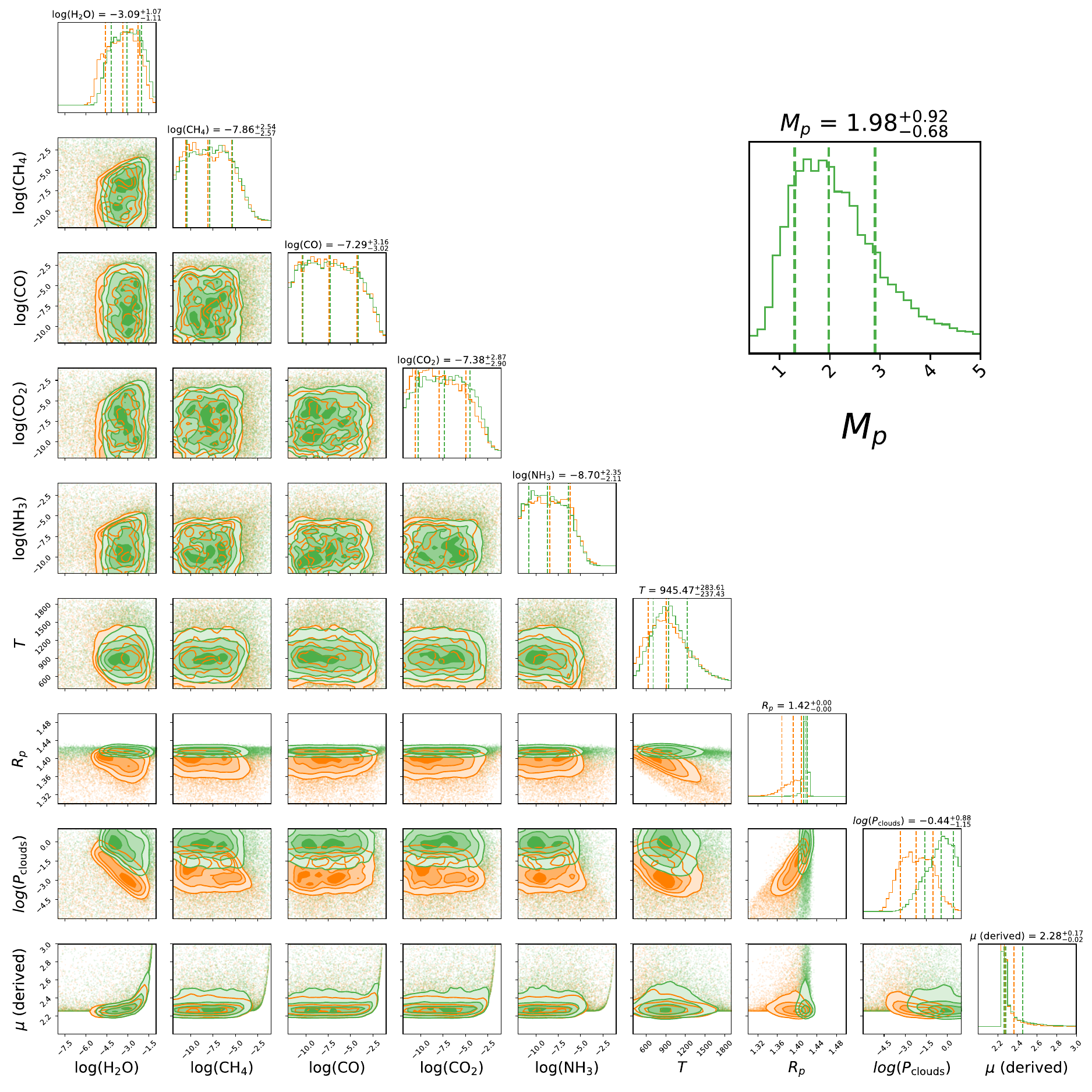}
    \caption{Hubble transit spectra (left) and posteriors distribution (right) for HD\,209458\,b  (\cite{Tsiaras_pop_study_trans}). Orange plots: the mass is known. Green plots: the mass is retrieved. Blue crosses:  Hubble observations.}
    \label{fig:hd209_ret}
\end{figure}

An adequate wavelength coverage is also very important to retrieve reliably the mass. To illustrate this point, we compare the results of Hubble observations for HD\,209458\,b (\cite{Tsiaras_pop_study_trans}) when the mass is known and retrieved with uniform priors.
The best fitted spectra and posterior distributions for both cases are presented in Figure\,\ref{fig:hd209_ret}.

Similarly to the high-altitude cloud case presented in \S 3.2, the retrieved trace gas abundances  and  temperature, while not being very precise,  are not affected by the mass uncertainties. The main differences appear in the retrieved radius and  cloud top pressure. The retrieved mass is not accurate: 1.98 $M_J$ instead of 0.73 $M_J$. The difference is significant (170\%) and demonstrates that a broad wavelength coverage and an adequate S/N is necessary to estimate correctly the mass through transit spectroscopy.

\subsection{Cloudy secondary atmospheres}

\begin{figure}[h]
\centering
    \includegraphics[align=c,width=0.5\textwidth]{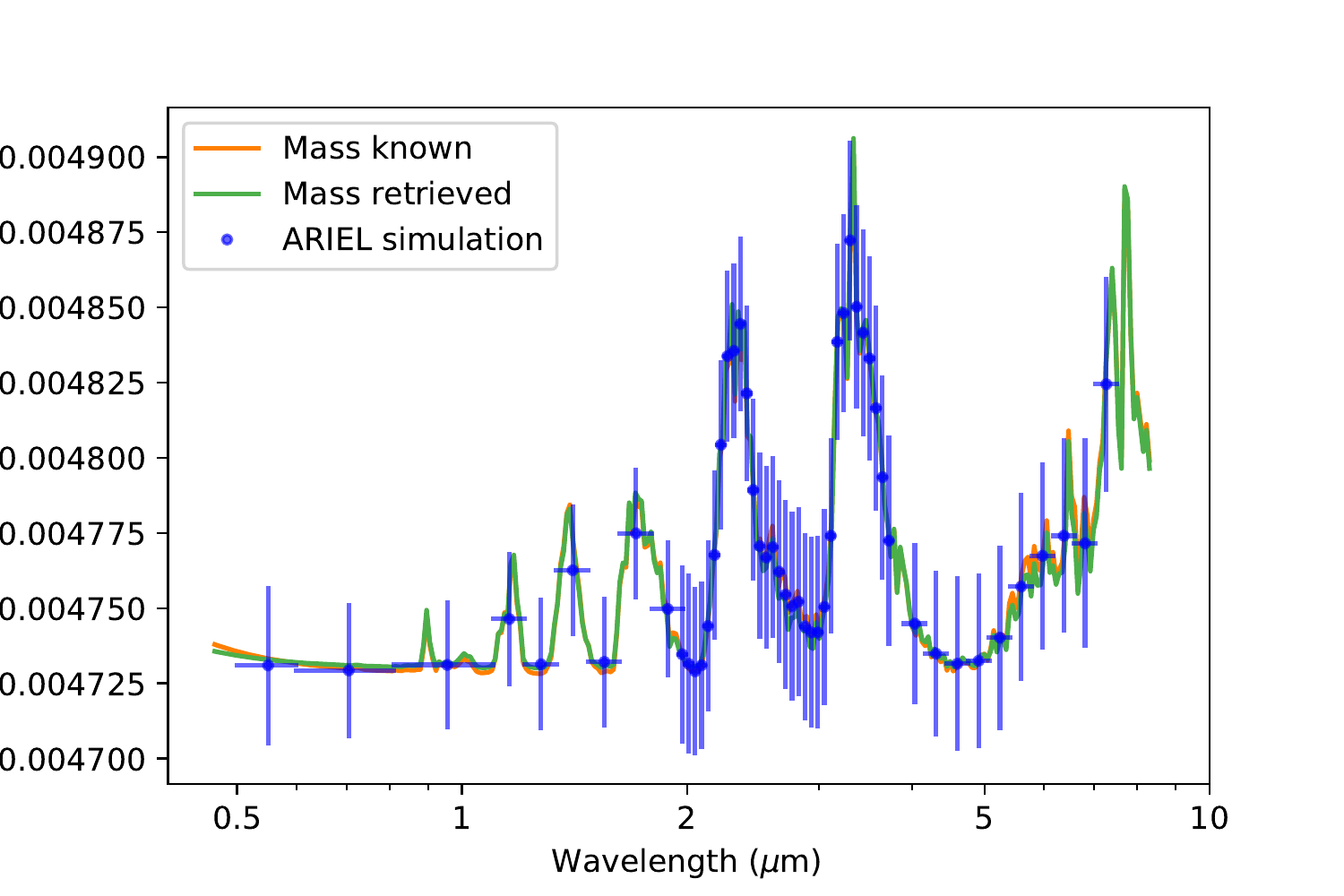}
    \includegraphics[align=c,width=0.49\textwidth]{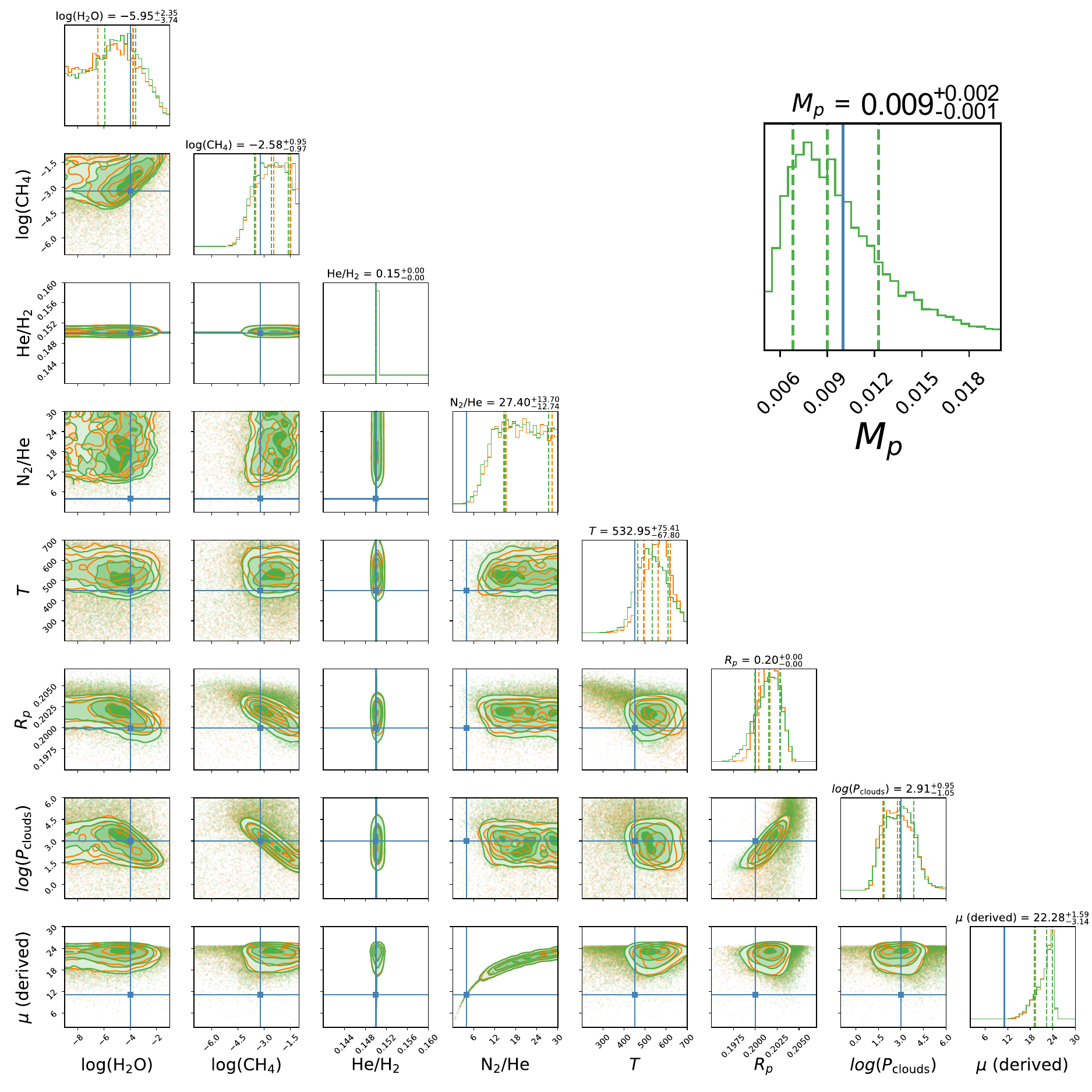}
    \includegraphics[align=c,width=0.5\textwidth]{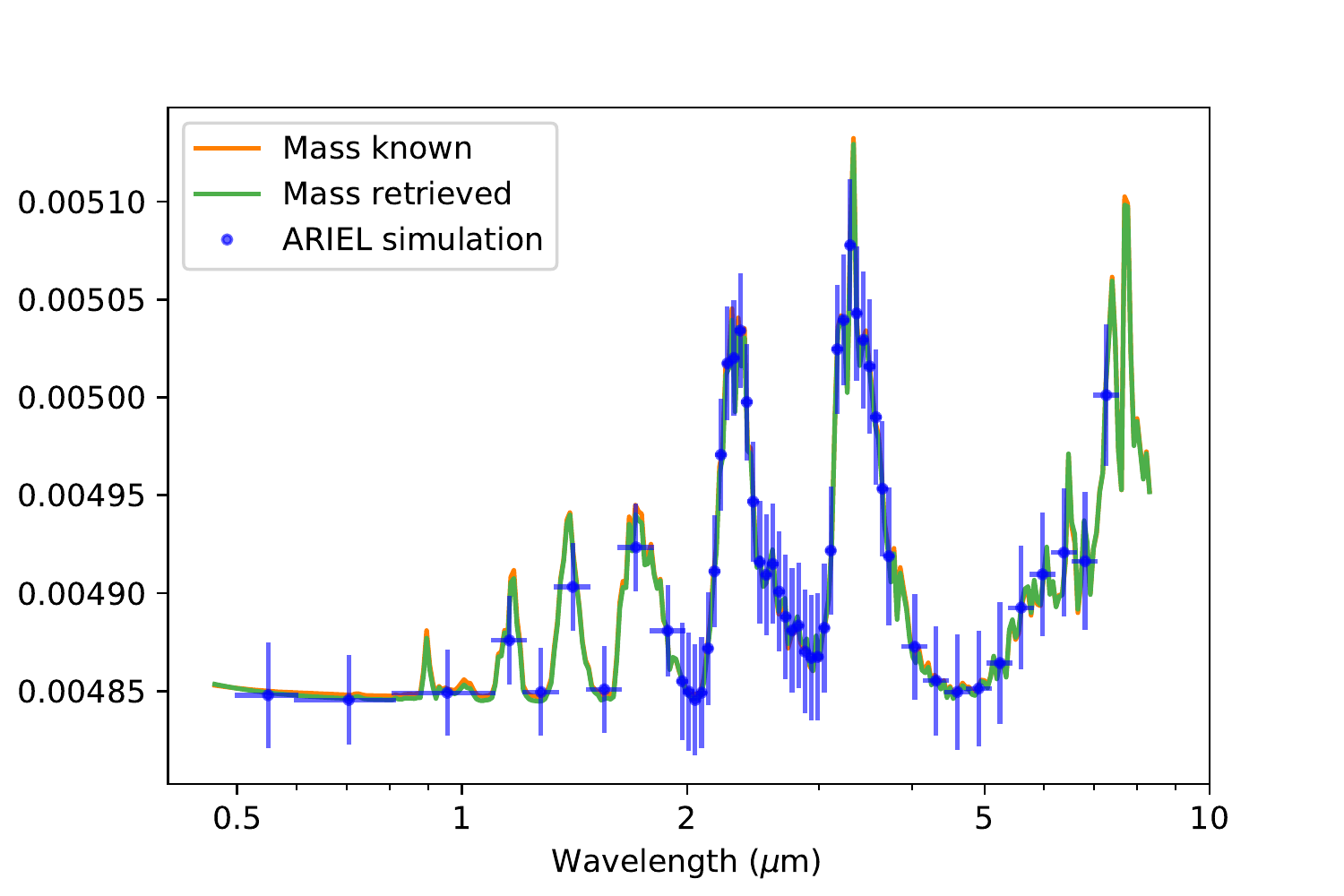}
    \includegraphics[align=c,width=0.49\textwidth]{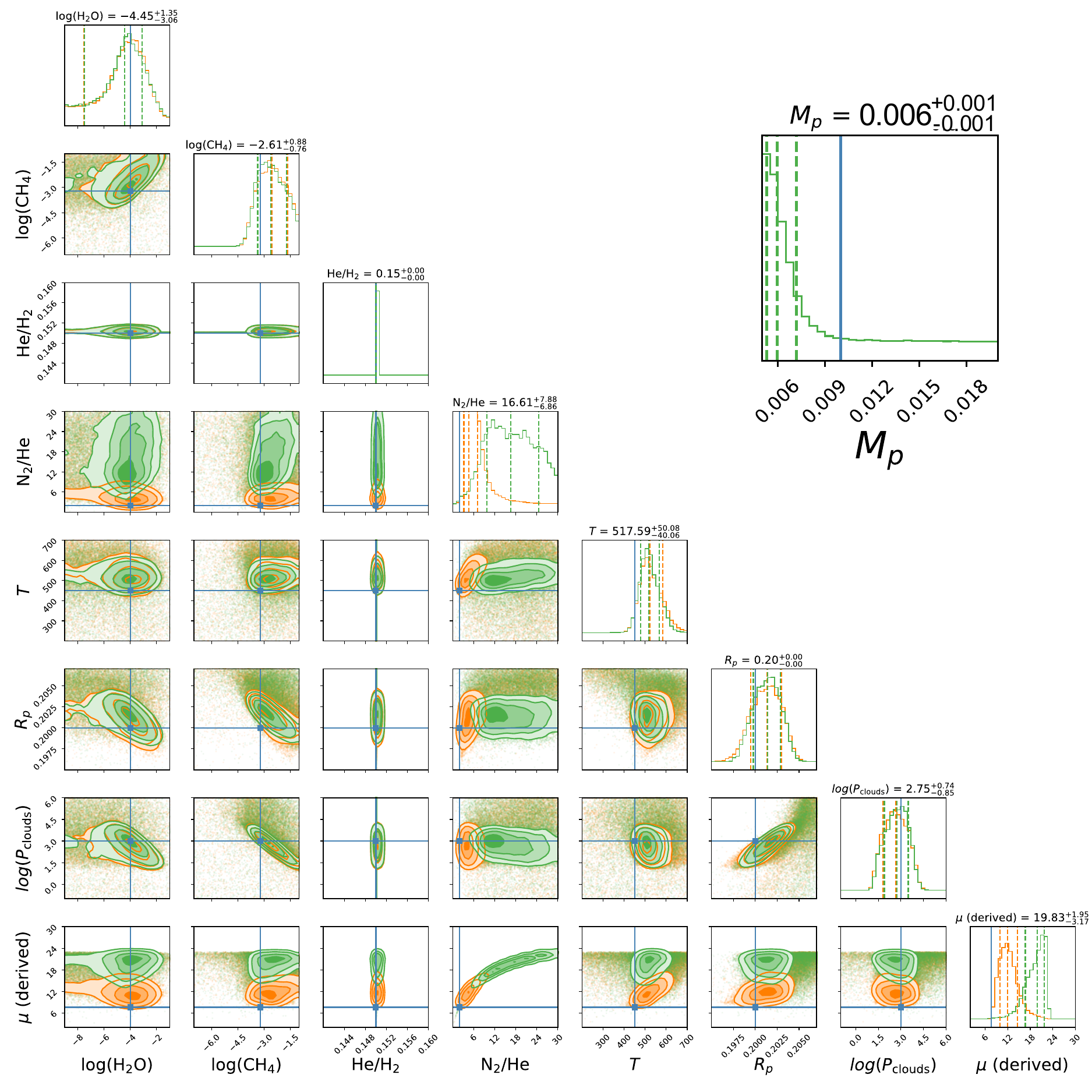}
    \caption{ARIEL simulated spectra (left) and posteriors distribution (right) for a planet with a  cloudy secondary atmosphere. The top cloud pressure is $10^{-2}$ bar. Top: $\mu = 11.1$, bottom: $\mu = 7.6$. Orange plots: the mass is known. Green plots: the mass is retrieved. Blue crosses:  simulated ARIEL observations obtained in one transit (left) and ground truth values (right).}
    \label{fig:spec_heavy_clouds}
\end{figure}

Finally, we investigate the case of cloudy secondary atmospheres. In this case, the mass is expected to be degenerate with both the mean molecular weight (see Figure \ref{fig:spec_heavy}) and the cloud top pressure (see Figure \ref{fig:post_clouds} and \ref{fig:forward_clouds}). In Figure \ref{fig:spec_heavy_clouds}, we show the simulated spectra and posteriors for two different mean molecular weights: $\mu = 11.1$ and $\mu = 7.6$ (corresponding $N_2/He$ ratios of 4 and 2);    opaque clouds are added at $10^{-2}$ bar.

In Figure \ref{fig:spec_heavy_clouds} we show  the cases $\mu = 11.1$ (top) and $\mu = 7.6$ (bottom). In the case $\mu = 7.6$,  the atmosphere is lighter and presents a better signal. Here prior knowledge of the mass allows  to break the degeneracy and retrieve the appropriate $N_2/He$, as well as the cloud pressure.  Without prior information about the mass, the model retrieves the trace gas abundances and the temperature with equal accuracy/precision but it is not able to constrain  the mean molecular weight.
In the case of $\mu = 11.1$, the retrieval does not properly constrain the $N_2/He$ ratio and provides a wrong lower limit on its value, leading to a biased estimate of the mean molecular weight $\mu$. Here, additional observations are needed to increase the S/N and to constrain the mean molecular weight. 

 From these examples, we deduce that the degeneracy with the mean molecular weight is more serious than the degeneracy with clouds, especially as the grey cloud assumption adopted here is pessimistic. Other more realistic cloud models (\cite{Lee_haze_model} or \cite{Madhu_retrieval_method}) would be more transparent at least in some spectral windows, so that information from the deeper atmosphere could be captured.

\section{Discussion and CONCLUSIONS}

Table \ref{table} summarises the cases investigated in our study, showing the mass uncertainties in percent and the parameters  affected if we ignore the planetary mass.

\begin{table}
\[ \begin{array}{|l|c|c|c|c|c|c|c|}
\mbox{Type of planet} & \mbox{HJ or SN, clear-sky} & \mbox{HJ, high opaque clouds}   &  \mbox{SE, clear-sky} & \mbox{SE, high opaque clouds}  &  \mbox{HJ HST} \\
\hline
\mbox{Mass uncertainty} & <10\% & 10 - 60\%     & <10\%^{\dag} & \mbox{degenerate} & 170\% \\
\mbox{Temperature} & \emptyset & \emptyset     &  \emptyset & \emptyset & \emptyset \\
\mbox{Trace composition} & \emptyset & \emptyset     &  \emptyset & \emptyset & \emptyset \\
\mbox{Main composition} & \emptyset & \emptyset     &  \mbox{yes} & \mbox{yes} & \emptyset \\
\mbox{Radius} &  \emptyset & \mbox{yes} &        \emptyset & \mbox{yes} & \mbox{yes} \\
\mbox{Clouds} &  NA & \mbox{yes} &        NA &   \mbox{yes} & \mbox{yes} \\
\end{array}\]
\caption{Note: HJ -- Hot Jupiter; SN -- Sub-Neptune with H/He-rich atmosphere; SE -- Super Earth with secondary atmosphere; Yes -- affected by the knowledge of the mass; $\emptyset$ -- Unaffected by the knowledge of the mass; NA -- Not Applicable. \newline 
$^{\dag}$ with adequate S/N and wavelength coverage. } \label{table}
\end{table}

For clear-sky, gaseous  atmospheres we find the same posterior distributions when the mass is known or retrieved. The retrieved mass is very accurate, with a precision of more than 10\%, provided the wavelength coverage and S/N are adequate. 

When opaque clouds are included in the simulations, the uncertainties in the retrieved radius and mass increase, especially for high altitude clouds. The error  in the retrieved mass  is up to 60\% for our worst case scenario, i.e.  a cloud pressure at $10^{-3}$ bar. Additionally, we find that the posterior distributions of the retrieved radius and mass are no longer  centred around their true values, indicating that solutions with different masses, radii and cloud parameters present similar likelihood. By contrast, atmospheric parameters such as the temperature and trace gas abundances appear to be unaffected by the knowledge of -- or lack of -- the mass. 

Secondary atmospheres are more challenging due to the higher degree of freedom for the atmospheric main component. For broad wavelength ranges and adequate S/N observations, the mass can still be retrieved accurately and precisely if clouds are not present, and so can all the other parameters.
We confirm the results in \cite{Batalha_mass} concerning secondary atmospheres dominated by multiple species, for which a degeneracy may exist. Here, prior information about the mass may help to extract the main constituent ratios. However, we also show that it is possible to retrieve the mass of full H$_2$ and full N$_2$ planets down to an accuracy of 10\% when the S/N  is sufficient. This confirms the results from \cite{deWit_mass}. 

 When clouds are added, we find that the mass uncertainties may impact substantially the retrieval of the mean molecular weight: an independent characterisation of the mass would therefore be helpful to capture/confirm the main constituents.

In the context of large scale surveys (ARIEL) and dedicated studies (JWST) of exoplanetary atmospheres, our results indicate that  constraining  the planetary mass for  secondary atmospheres is important to ensure that we fully exploit the information content of the spectra. Current mass estimates found in exoplanet databases, which are mainly coming from  radial velocity follow-up confirmations, have typical error bars  of the order of 10\%.
Such small uncertainties guarantee an excellent prior knowledge for the mass in retrieval simulations, even for overcast planets. 

Planets smaller than Neptune have larger mass errors, often larger than 50\%. This uncertainty 
may contribute to the degeneracy in retrieving the mean molecular weight of the atmosphere, especially when clouds are present. Radial velocity campaigns should therefore prioritise the mass characterisation of low-gravity planets as in the other cases, transit spectroscopy retrievals appear to be sufficiently robust to mass uncertainties.

\vspace{5mm}
\noindent\textbf{Acknowledgements}

This project has received funding from the European Research Council (ERC) under the European Union's Horizon 2020 research and innovation programme (grant agreement No 758892, ExoAI; No. 776403, ExoplANETS A) and under the European Union's Seventh Framework Programme (FP7/2007-2013)/ ERC grant agreement numbers 617119 (ExoLights). Furthermore, we acknowledge funding by the Science and Technology Funding Council (STFC) grants: ST/K502406/1, ST/P000282/1, ST/P002153/1, ST/T001836/1 and ST/S002634/1. We also thank the referee for his/her relevant comments that have greatly improved the quality of this manuscript.




\newpage
\bibliographystyle{aasjournal}
\bibliography{main}


\renewcommand{\floatpagefraction}{.9}%

\appendix

\section*{Star-Planet parameters used for this study} \label{star_planet}


\[ \begin{array}{|l|c|c|c|}
\mbox{Parameters} & Hot\ Jupiter & Super\ Earth  \\
\hline
\mbox{$R_s (R_{sun})$} & 1.19 & 0.3 \\
\mbox{$T_s (K)$} & 6091 & 3671 \\
\mbox{$M_s (M_{sun})$} & 1.23 & 0.4\\
\mbox{$Distance (pc)$} & 48 & 2.6\\
\mbox{$R_p (R_{Jupiter})$} & 1.39 & 0.2  \\
\mbox{$M_p (M_{Jupiter})$} & 0.73 & 0.01 \\
\mbox{$T_p (K)$} & 1450 & 450
\end{array}\]
\\

\section*{Derivation of transit equation}\label{Appendix_transit}
\begin{figure}[h]
\begin{center}
\begin{tikzpicture}

\draw [fill= black, fill opacity=0.1] (0,0) circle (4);
\draw [fill= black, fill opacity=0.3] (0,0) circle (2.8);
\draw [] (0,0) circle (3.5);
\draw [->] (-5,3.5) -- (5,3.5);

\draw [-] (0,0) -- (0,3.5);
\draw [-] (0,0) -- (1.5,3.5);

\draw [dashed, <->] (-0.1,-0.1) -- (-1.8,-1.8);
\node (a) at (-1.2, -0.8) {$R_0$};
\draw [dashed, <->] (0,3.7) -- (1.5,3.7);
\node (b) at (0.5, 3.85) {$x$};
\draw [dashed, <->] (-0.2,2.9) -- (-0.2,3.4);
\node (c) at (-0.5, 3.1) {$z$};

\draw [dashed, <->] (1.5,3.2) -- (1.65,3.5);
\node (d) at (1.8, 3.3) {$z'$};

\draw [dashed, <->] (-0.8,0) -- (-0.8,3.4);
\node (e) at (-1, 1.5) {$r$};

\draw [dashed, <->] (0.5,-0.3) -- (2.1,3.2);
\node (f) at (1, 1.5) {$r'$};

\draw (6,3.5+0.5) -- (7,3.5+0) -- (6,3.5-0.5);
\draw (6.4,3.5+0.3) -- (6.4,3.5-0.3);

\end{tikzpicture}
\end{center}
\caption{Illustration of the transmission of the stellar radiation through an exoplanet atmosphere during a transit event.  $R_0$ is the  radius at which when the planet becomes fully opaque in absence of clouds.  For a given point in the atmosphere, $z$ is the altitude normal to the sun-observer connecting line and $x$ is the projected distance from that normal to the point. $r'$ is the distance from the point to the planetary centre. In addition, we define $r = R_0 + z$ and $z' = r' - r$.} \label{fig:geometry}
\end{figure}
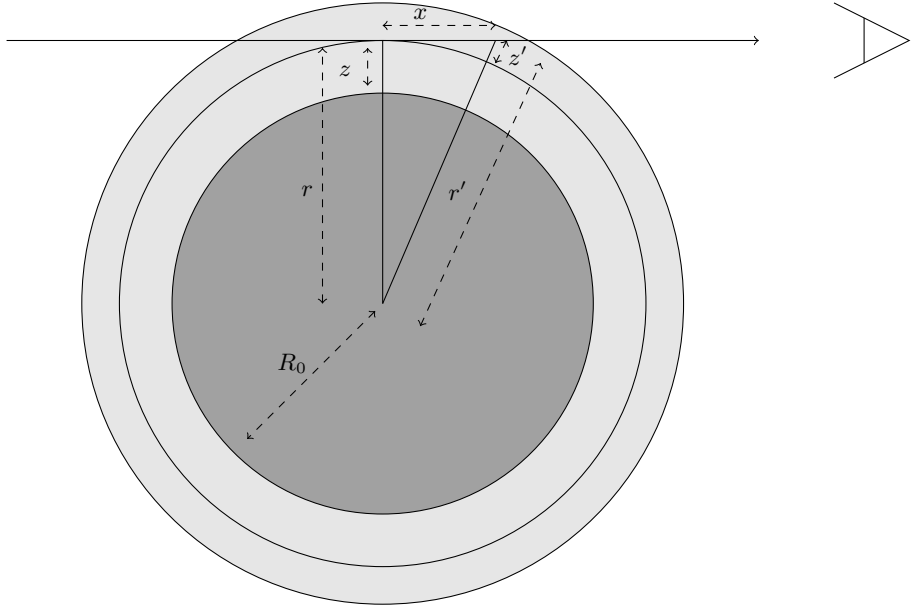

The transit geometry and relevant variables are illustrated in Figure \ref{fig:geometry}.
The normalised differential flux $\Delta$ between in-transit $F_{in}$ and out-transit $F_{out}$ can be calculated as:
\begin{equation}
    \Delta = \frac{F_{out} - F_{in}}{F_{out}}=\frac{R_p(\lambda)^2}{R_s^2},
\end{equation}
where $R_p(\lambda)$ is the wavelength dependent radius which includes the atmospheric contribution  and $R_s$ is the stellar radius. \\ 
The wavelength-dependent contribution of the atmosphere starts at $R_0$, we have:

\begin{equation}\label{eq:contrib_atm}
    \pi R_p(\lambda)^2 = C_{surf} + C_{atm}(\lambda) = 2\pi \int_0^{R_0}r \text{d}r + 2 \pi \int_{R_0}^{\infty} r(1-e^{-\tau(r,\lambda)})\text{d}r,
\end{equation}
where we have introduced the optical depth $\tau(r,\lambda)$. \\
The optical depth represents the atmospheric absorption at a given altitude  integrated along the line of sight:
\begin{equation}
    \tau(r,\lambda) = 2 \int_0^{x_f}\sum_i n_i(r')\sigma_i(r',\lambda) \text{d}x,
\end{equation}
where $n_i$ is the number density of the $i^{th}$ species and $\sigma_i$ is the cross section of the $i^{th}$ species.   $x_f$ is the maximum distance in the atmospheric layer along the line of sight. \\ 
For the rest of this derivation, we consider standard assumptions in current retrieval models published in the literature \citep{Waldmann_taurex1, Waldmann_taurex2, irwin2008, chimera, Ormel_arcis, harrington_bart, Mollire_petitrad, Kitzmann, Lavie_helios, MacDonald_hd209, Madhu_retrieval_method, Gandhi_retrieval, benneke_retrieval, Zhang_platon, cubillos_pirat, alrefaie2019taurex}. We assume the atmosphere is isothermal and in hydrostatic equilibrium. The scale height $H$ is therefore defined as:
\begin{equation}
    H = \frac{k_b T (R_0+z)^2}{\mu M_p G}
\end{equation}
 where $k_b$ the Boltzmann constant, $T$ the temperature and $\mu$ the mean molecular weight and $G$ the gravitational constant.
One can write the the number density as:
\begin{equation}
    n_i(r') = n_i(R_0)e^{- \frac{r'-R_0}{H}} = n_{0i} e^{- \frac{z+z'}{H}},
\end{equation}
For  simplicity we write $n_i(R_0) = n_{0i}$. 
\\
Using Pythagoras' theorem and neglecting second order terms of $z'$:
\begin{equation}
    (R_0+z)^2 +x^2 = (R_0+z+z')^2,
\end{equation}
\begin{equation}
    z' = \frac{x^2}{2(R_0 + z)},
\end{equation}
We can now rewrite $\tau$ as: 
\begin{equation}
    \tau(z,\lambda) = 2 \int_0^{x_f}\sum_i n_{0i}e^{-\frac{z}{H}}e^{-\frac{x^2}{2(R_0+z)H}}\sigma_i(p,T,\lambda) \text{d}x.
\end{equation}
Finally, the  contribution of the entire atmosphere can be estimated as:
\begin{equation}
    C_{atm}(\lambda) = 2 \pi \int_{R_0}^{\infty} (R_0 + z) \left(1-\exp\left [ -2 \int_0^{x_f}\sum_i n_{0i}e^{-\frac{z}{H}}e^{-\frac{x^2}{2(R_0+z)H}}\sigma_i(p,T,\lambda ) \text{d}x \right ]\right) \text{d}z.
\end{equation}
The planetary mass  appears only in the exponent, influencing solely the optical depth $\tau$.  We use here the same assumptions made in the retrieval analysis to estimate the cross sections $\sigma_i$. We note that $\sigma_i$  depend only on   temperature and  pressure: $\sigma_i(r',\lambda ) = \sigma_i(p(r'),T(r'),\lambda )$. Considering $\sigma_i(p,T,\lambda)$, we approximate their values by  using a linear interpolation with respect to  $p$ and $T$.
We assume the temperature of the planet is isothermal and will not change dramatically with variation of the mass. Therefore, the atmospheric temperature can be interpolated between two known values from cross-section tables (for example: ExoMol, ExoTransmit \cite{Kempton_2017_exotransmit}, HITEMP or HITRAN). The pressure, however, needs to be integrated along the  line of sight and we expect large variations: we therefore use a list of values labeled $p_j$ and interpolate $\sigma_i$ between two of these values.  We have:
\begin{equation}
    \sigma_i(T) = \sigma_i(T_1)+\frac{\sigma_i(T_2)-\sigma_i(T_1)}{T_2-T_1}(T-T_1),
\end{equation}
\begin{equation}
    \sigma_i(p) = \sigma_i(p_j)+\frac{\sigma_i(p_{j+1})-\sigma_i(p_j)}{p_{j+1}-p_j}(p-p_j),
\end{equation}
where $T_1$, $T_2$, $p_j$ and $p_{j+1}$ are fixed temperatures and pressures. Since the pressure differences across the $x$ axis are large --larger than the interpolation intervals--, we sum over intervals $(x_j, x_{j+1})$ of known pressures ($p_j, p_{j+1}$).
We define: 
\begin{equation}
    K_{ij}(T) = \frac{ \sigma_i(p_{j+1}, T, \lambda)-\sigma_i(p_j, T, \lambda)}{p_{j+1}-P_j},
\end{equation}
and estimate $\sigma |^{x_{j+1}}_{x_j}$ in the interval $(x_j, x_{j+1})$:
\begin{equation}
    \sigma_i(p,T,\lambda)|^{x_{j+1}}_{x_j} = \sigma_i(p_j, \lambda)+K_{ij}(T, \lambda)(p-p_j)
    \label{1}
\end{equation}
We include eq. \ref{1} in the expression of $\tau$, we get:
\begin{equation}
    \tau(z,\lambda) =\sum_j  \int_{x_j}^{x_{j+1}}\sum_i n_{0i}e^{-\frac{z}{H}}e^{-\frac{x^2}{2(R_0+z)H}}(\sigma_i(p_j,T,\lambda) - K_{ij}(T,\lambda) p_j +K_{ij}(T,\lambda) p )dx.
\end{equation}\label{master}
Knowing that the atmospheric pressure as a function of $x, z$ is $p = p_0 e^{-\frac{z}{H}}e^{-\frac{x^2}{2(R_0+z)H}}$, we can now calculate the integral:
\begin{equation}
    \int_{x_j}^{x_{j+1}} e^{-\frac{x^2}{L}} = \frac{1}{2}\sqrt{\pi L}\left[  \text{erf} (\frac{x}{\sqrt{L}}) \right]_{x_j}^{x_{j+1}},
\end{equation}
where $L$ is a normalization constant. Concerning the integration boundaries, we have $x_j = x(p_j)$ which translates into:
\begin{equation}
    x_j = \sqrt{-2(R_0 + z)H\left[ \ln \left ( \frac{p_j}{p_0} \right )+\frac{z}{H}\right]}
\end{equation}
Defining for convenience:
\begin{equation}
    I_j(z) = \text{erf} \left(\sqrt{- (\ln (\frac{p_{j+1}}{p_0})+\frac{z}{H})} \right) - \text{erf} \left(\sqrt{- (ln(\frac{p_{j}}{p_0})+\frac{z}{H})} \right),
\end{equation}
and:
\begin{equation}
    I_j'(z) = \text{erf} \left(\sqrt{-2 (ln(\frac{p_{j+1}}{p_0})+\frac{z}{H})} \right) - \text{erf} \left(\sqrt{-2 (\ln (\frac{p_{j}}{p_0})+\frac{z}{H})} \right),
\end{equation}
we get:
\begin{equation}
    \int_{x_j}^{x_{j+1}} e^{-\frac{x^2}{2(R_0+z)H}} = \frac{1}{2}\sqrt{2 \pi (R_0+z)H } I_j(z),
\end{equation}
and:
\begin{equation}
    \int_{x_j}^{x_{j+1}} e^{-\frac{x^2}{(R_0+z)H}} = \frac{1}{2}\sqrt{ \pi (R_0+z)H } I_j'(z),
\end{equation}
This result leads to:
\begin{equation}
    \tau(z,\lambda) = \sum_j \sum_i n_{0i}e^{-\frac{z}{H}} \sqrt{\pi (R_0 +z)H} \left((\sigma_i(p_j,T)-K_{ij} p_j)\sqrt{2} I_j + K_{ij} e^{-\frac{z}{H}}p_0 I_j' \right)
\end{equation}
Now the temperature dependence of $\sigma$ can be added in the same way. Here we assume the atmosphere to be isothermal, so we do not have to consider the temperature outside our reference $\sigma(T_2)$ and $\sigma(T_1)$. This leads to:
\begin{equation}\label{eq:tau}
    \tau = \sum_j \sum_i n_{0i} e^{-\frac{z}{H}}\sqrt{\pi (R_0+z) H} \left( \sqrt{2} I_j\sigma_i(p_j,T_1) +K^{T}_{ij} \sqrt{2} I_j(T-T_1) + (K^{p}_{ij} + K^{X}_{ij} (T-T_1))(p_0 I_j' e^{-\frac{z}{H}} - \sqrt{2} I_j p_j) \right),
\end{equation}
where the coefficients  $K^{p}_{ij}$, $K^{T}_{ij}$ and $K^{X}_{ij}$ are only wavelength dependent and can be calculated from tables:
\begin{equation}
    K^{p}_{ij} = \frac{\sigma_i(p_{j+1},T_1)-\sigma_i(p_j,T_1)}{p_{j+1}-p_j},
\end{equation}
\begin{equation}
    K^{T}_{ij} = \frac{\sigma_i(p_j,T_2)-\sigma_i(p_j,T_1)}{T_2-T_1},
\end{equation}
\begin{equation}
    K^{X}_{ij} = \frac{1}{p_{j+1}-p_j}\frac{1}{T_2-T_1}\left( \sigma_i(p_{j+1},T_2) - \sigma_i(p_{j+1},T_1) - \sigma_i(p_j,T_2) + \sigma_i(p_j,T_1)\right),
\end{equation}
We replace:
\begin{equation}
    H = \frac{k_b T (R_0 + z)^2}{\mu M_p G},
\end{equation}
By considering the cross sections  constant with pressure and temperature, $p_{j+1} = 0$ and $p_{j} = p_0 e^{-\frac{z}{H}}$  the equations for $I_j = 1$ and $\tau$ are simplified and we finally get:
\begin{equation}
    \tau = \sum_i n_{0i} \sigma_i(p_0,T_0) e^{-\frac{z}{H}}\sqrt{2 \pi (R_0+z) H} .
\end{equation}
  
 We can investigate the contribution of grey clouds  by separating the atmospheric terms  below and above clouds in equation \ref{eq:contrib_atm}, we obtain:

\begin{equation}
    \pi R_p(\lambda)^2 = C_{surf} + C_{clouds} + C_{atm}(\lambda), = 2\pi \int_0^{R_0}r \text{d}r + 2\pi \int_{0}^{z_c}(R_0+z) \text{d}z + 2 \pi \int_{z_c}^{\infty} (R_0+z)(1-e^{-\tau(z,\lambda)})\text{d}z,
    \label{eq:cloud}
\end{equation}
where $\tau(z,\lambda)$ is given in equation \ref{eq:tau} and:
\begin{equation}
    z_c = - H\, \text{ln}(\frac{P_{c}}{P_{s}}),
\end{equation}
$P_{c}$  is the cloud top pressure and $P_{s}$ the pressure at the reference radius $R_0$ (10 bar in this paper).


\end{document}